\documentclass[preprint]{aastex}
\usepackage{amssymb, amsmath}
\usepackage{enumerate}
\DeclareMathOperator{\sech}{sech}

\begin{document}

\title{A Data Cube Extraction Pipeline for a\\Coronagraphic Integral Field Spectrograph}

\author{Neil Zimmerman\altaffilmark{1,2},
Douglas Brenner\altaffilmark{2},
Ben R. Oppenheimer\altaffilmark{2,1},
Ian R. Parry\altaffilmark{3},
Sasha Hinkley\altaffilmark{4,5},
Stephanie Hunt\altaffilmark{3},
and Robin Roberts\altaffilmark{2}}

\altaffiltext{1}{Department of Astronomy, Columbia University, 550 W 120$^{\mbox{\tiny{th}}}$ St, New York, NY 10027; e-mail address: \email{neil@astro.columbia.edu}}
\altaffiltext{2}{Astrophysics Department, American Museum of Natural History, Central Park West at W 79$^{\mbox{\tiny{th}}}$ St, New York, NY 10024}
\altaffiltext{3}{Institute of Astronomy, University of Cambridge, Madingley Road, Cambridge CB3 OHA, UK}
\altaffiltext{4}{Department of Astronomy, California Institute of Technology, 1200 E. California Blvd., MC 249-17, Pasadena, CA 91125} 
\altaffiltext{5}{Sagan Fellow}

\slugcomment{To appear in the June 2011 issue of PASP.}

\begin{abstract}

Project 1640 is a high contrast near-infrared instrument probing the vicinities
of nearby stars through the unique combination of an integral field
spectrograph with a Lyot coronagraph and a high-order adaptive optics system.
The extraordinary data reduction demands, similar those which several new
exoplanet imaging instruments will face in the near future, have been met by
the novel software algorithms described herein. The Project 1640 Data Cube
Extraction Pipeline (PCXP) automates the translation of $3.8\times10^4$ closely
packed, coarsely sampled spectra to a data cube. We implement a robust
empirical model of the spectrograph focal plane geometry to register the
detector image at sub-pixel precision, and map the cube extraction. We
demonstrate our ability to accurately retrieve source spectra based on an
observation of Saturn's moon Titan.

\end{abstract}


\section{Introduction}

In recent years an assortment of new astronomical techniques have evolved to
address the challenges of imaging faint objects and disk structure at close
angular separations to nearby stars. A major scientific motivation for these
efforts is the direct detection and characterization of low-mass companion
bodies orbiting at separations between $\sim$ 5 and 100 AU. These objects are
beyond the reach of conventional optical imaging due to the extreme contrast in
brightness with respect to the primary star. In such cases, even under ideal
observing conditions the diffracted light of the primary star overwhelms the
neighboring source of interest. The various methods of manipulating a star's
light to enable investigation of its immediate environment are collectively
referred to as high-contrast imaging. For a recent review of this field,
see~\cite{OppHinkley09}. The acquisition of spectra of young, sub-stellar mass
objects in this newly opened parameter space will ultimately lead to a
breakthrough in our understanding of exoplanet
populations~\citep{BeichmanPASP}.

\begin{figure}[tb]
\epsscale{1.0}
\plotone{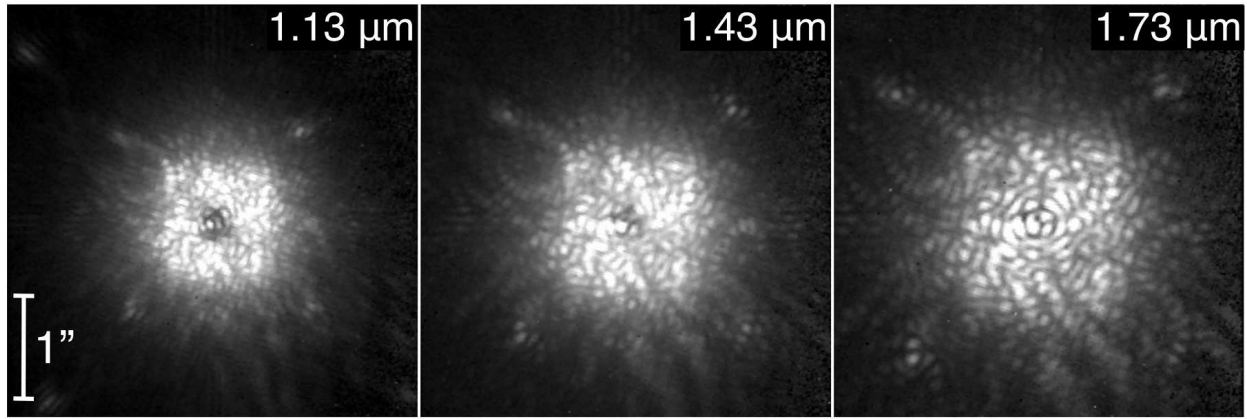}

\caption{Three of the 23 channels making up an example P1640 data cube, formed
from a 154 s exposure of the coronagraphically occulted star HD 27946 ($V$ =
5.3). The three channel images shown here, each consisting of $200\times200$
\textit{spaxels}, are displayed with a square root stretch. A halo of
well-defined speckles surround the focal plane mask, expanding with wavelength.
The cube configuration of the data enables post-processing algorithms to take
advantage of the chromaticity of speckles to reduce their influence on
detection limits.} 

%\caption{Three of the 23 channels making up an example P1640 data cube, formed
%from a 154 s exposure of the coronagraphically occulted star Regulus ($V$ =
%1.35). A halo of well-defined speckles surround the focal plane mask, expanding
%with wavelength.  The cube configuration of the data enables post-processing
%algorithms to take advantage of the chromaticity of speckles to reduce their
%influence on detection limits.} 

\label{fig:examplecube}
\end{figure}

Project 1640 (P1640) is the first of several instruments to approach the high
contrast imaging problem through a combination of high-order adaptive optics, a
Lyot coronagraph, and an integral field spectrograph~\citep{p1640}.
Forthcoming instruments using a similar design include the Gemini Planet Imager
(GPI)~\citep{MacintoshSPIE}, the Very Large Telescope Spectro-Polarimetric
High-contrast Exoplanet REsearch (VLT-SPHERE) project~\citep{Beuzit}, and the
Subaru Telescope Planetary Origins Imaging Spectrograph (POISE)~\citep{poise}.
While previous efforts have used integral field spectrographs for high contrast
imaging~\citep[e.g.][]{thatte, mcelwain, janson}, and Lyot coronagraphs have also been
employed for surveys of nearby stars~\citep[e.g.][]{chauvin, leconte}, P1640 is
the first instrument to combine these two technologies. The coronagraph
component, based on the Fourier optics concept described in~\cite{anandlyot},
rejects the core of the target star's point spread function (PSF) and
attenuates the surrounding diffraction rings. Provided that the adaptive optics
(AO) system upstream of the coronagraph has corrected the star's PSF to near
the diffraction limit, then the dominant source of noise in the image exiting
the coronagraph takes the form of a halo of speckles surrounding the occulted
star, as in Figure~\ref{fig:examplecube}~\citep{racine, highstrehl}. These
relatively long-lived, point source-like artifacts are caused by uncorrected
wave front aberrations, and limit the dynamic range of the data unless further
processing is carried out~\citep{dynamicrange}.

The integral field spectrograph, also referred to as the integral field unit
(IFU), is situated after the coronagraph and provides spatially resolved
spectra for a grid of points across the field of view~\citep{bacon}. The
reduced form of data acquired with an IFU is a stack of simultaneous narrowband
images spanning the instrument's wavelength range, often called a data cube. An
example of part of a P1640 data cube is shown in Figure~\ref{fig:examplecube}.
One benefit the IFU provides is enabling the observer to measure the spectrum
of any source at any position in the field of view. This is not possible with a
conventional spectrograph, which can only use one spatial dimension at a time
to discriminate against other sources in the field of view. The second purpose
of the IFU is to exploit the chromatic behavior of the
speckles~\citep{sparksandford}. Speckles are optical in origin and their
separation from the target star is linearly proportional to wavelength, at
least to a first-order approximation (see Figure~\ref{fig:examplecube}). By
comparing the images in different channels of the data cube, the observer can
discriminate the speckles from a true point source, whose position should
remain constant with wavelength. Furthermore, automated post-processing
software can use the chromaticity of the speckles to subtract a large component
of them from the data, taking a reduced data cube as input and generating a
speckle-suppressed version~\citep{creppSSP, pueyospeckle}.

The unusual properties of data generated by the P1640 IFU necessitate novel
reduction techniques to reach the point where inspection, spectrophotometry,
astrometry, and advanced post-processing techniques like speckle suppression
can begin. In the scope of this article, we describe the software created to
translate rapidly the raw data from the IFU camera to a set of data cubes ready
for further analysis.

\section{Project 1640 Design and Data Acquisition}
\label{sec:design}

During operation at Palomar Observatory, P1640 receives a wave front-corrected
beam of the target star's light from the 200" Hale Telescope AO system. The
current AO system, the 241-actuator PALAO~\citep{palao}, will soon be upgraded
to the 3388-actuator PALM-3000~\citep{palm3k}. Upon entering the instrument,
the light passes through an apodized Lyot coronagraph, followed by an integral
field spectrograph, which contains a near-infrared camera.

In addition to the focal plane mask and Lyot stop of a traditional Lyot
coronagraph, P1640 uses a pupil plane apodization mask to optimize the
starlight suppression based on the telescope pupil shape~\citep{apodizer}. The
beam exiting the coronagraph comes to a focus on a $200\times200$ square
microlens array at the entrance of the spectrograph. Immediately after the
microlens array, the light is collimated to form a pupil on a wedge-shaped
prism, which disperses the light over the 1.1-1.8 $\mu$m wavelength range of
operation spanning the $J$ and $H$ bands. Additional optics focus the
$4\times10^4$ resulting spectra onto a Teledyne HAWAII-2 $2048\times2048$
pixel, HgCdTe, near-infrared detector. The field of view of the final image,
designed to match the control radius of the PALM-3000 AO system, is
$4\arcsec\times4\arcsec$. For further details on the optical and mechanical
design, see~\cite{p1640}.

For each exposure, the camera controller performs a sequence of non-destructive
reads on the detector array. In other words, the digitized value of each pixel
is periodically sampled while its voltage escalates. This technique, known as
up-the-ramp sampling, can result in the read noise being reduced by a factor of
$\sqrt{N/12}$ in a reduced image when the counts versus read slope is fit for
the $N$ samples of each pixel~\citep{uptheramp}. Up-the-ramp sampling also adds
an advantageous temporal dimension to our data. Speckle suppression algorithms
work best when the positions of the speckles are well-defined. For bright stars
with high signal-to-noise in individual read differences, it may be helpful to
``freeze'' the speckle pattern with the higher time resolution enabled in a
read-by-read data reduction. Our pipeline reduces the detector data with both
approaches: the non-destructive read (NDR) slope fit and consecutive read
differences.

The read sample interval is fixed at 7.7 seconds by the camera controller. The
sequence of reads are stored in a binary file containing the arrays of 16-bit
unsigned integer samples, which we refer to as a \textit{dat} file. The camera
controller also generates a separate FITS file with a header containing the
information about the telescope and instrument status, the target (coordinates,
magnitude, parallax, etc.), and the name of the \textit{dat} file corresponding to the
exposure. A typical observation of a Project 1640 science target is made up of
15 exposures, each containing 20 reads, giving a cumulative exposure time of
$38.6$ minutes. The resulting volume of raw data is 160 Mbytes for each
exposure's \textit{dat} file and 2.4 Gbytes in total.

\begin{figure}[tb]
\epsscale{1.0}
\plotone{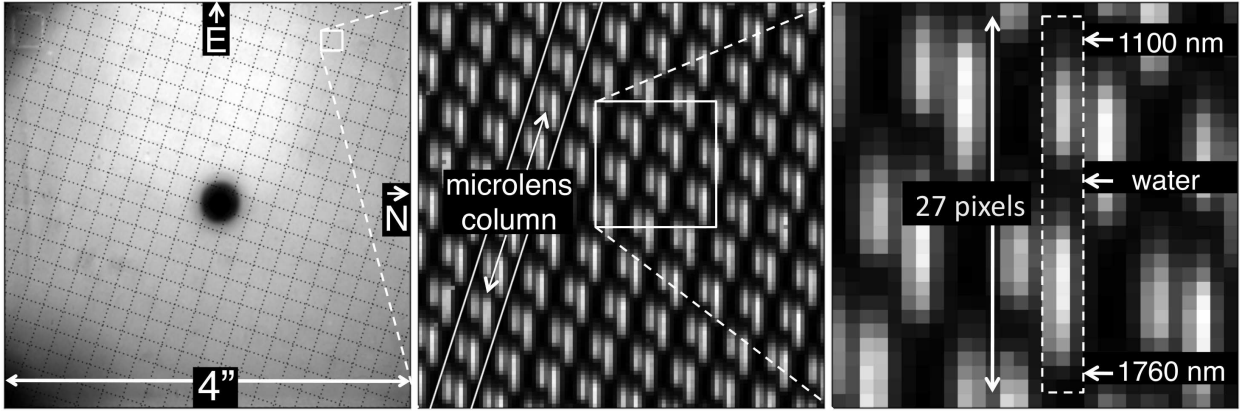}

\caption{A hierarchical diagram of the P1640 integral field spectrograph focal
plane. \textbf{Left}: an average of ten calibration Moon exposures, recorded by
the near-infrared detector at the spectrograph focal plane. The projection of
the IFU microlens array onto the detector is represented by a superimposed
dotted grid, with each gap spanning ten microlenses. \textbf{Middle}: expanding
a $100\times100$ pixel section reveals the underlying pattern of microlens
spectra.  The tilt of the square microlens array with respect to the detector
($-18.5^{\circ}$) interleaves the adjacent rows of spectra for an efficient use
of detector area.  \textbf{Right}: each spectrum, spanning $J$ and $H$ band
(1.1~$\mu$m to 1.8~$\mu$m), displays the prominent telluric water absorption
trough centered near 1.4~$\mu$m.} 

\label{fig:focpldiagram}
\end{figure}

The structure of the P1640 IFU focal plane, illuminated by Moonlight, is
depicted in Figure~\ref{fig:focpldiagram}. The microlens array, represented by
the dotted grid superimposed on the left panel, is rotated with respect to the
detector. This configuration interleaves the adjacent rows of microlens
spectra, thereby maximizing the efficiency of focal plane area usage. Along a
given column of microlenses, the mean interval between neighboring spectra is
3.3 pixels in the horizontal direction and 10.0 pixels in the vertical
direction. Each spectrum takes up a length of approximately 27 pixels in the
dispersion direction. 

\section{Spectrograph Focal Plane Model}
\label{sec:focplmodel}

Rather than relying purely on design predictions, we have written procedures to
empirically determine the IFU response, capturing the minute optical
distortions and alignment changes unique to each observing run. Two forms of
calibration data are used as input for the focal plane model. The first kind
are the spectrograph images formed by illuminating the instrument pupil with a
tunable laser source. These allow us to characterize the response at fixed
wavelengths across the passband. Secondly, during each observing run we observe
a broadband source of nearly uniform brightness across our field of
view---either the Moon or the twilight sky. In the case of the Moon calibration
images, the telescope AO correction loop was turned off, and several exposures
with different pointings were averaged. These images constrain the geometry of
the focal plane, including the positions and shapes of the individual spectra
formed on the detector, as well as large-scale variations in sensitivity across
the field of view due to vignetting.

\subsection{Spectrograph Point Spread Function Model}
\label{sec:psfmodel}

An accurate model of the monochromatic spectrograph point spread function (PSF)
is at the core of the IFU focal plane model. We emphasize the distinction here
from the coronagraph PSF, which is formed on the mircolens array at the
entrance of the IFU. The spectrograph PSF, on the contrary, is the signal
formed on the IFU focal plane from monochromatic light incident on an
individual microlens. In a laboratory environment before the first scientific
observing run, we illuminated the IFU with a tunable laser source and recorded
narrowband emission (bandwidth $<$ 4 nm) images at wavelengths in 0.01 $\mu$m
increments spanning the 0.7 $\mu$m operating band of the instrument. Any given
laser image shows a grid of $3.8\times10^4$ point spread functions, each
corresponding to a microlens illuminated by the beam entering the spectrograph.
From these images, we derived an analytic model of the spectrograph PSF
specific to the recorded wavelength, as follows: First, a script looped through
all of the PSFs in the spectrograph image, forming a 9$\times$9 pixel
\emph{mean} PSF based on the subset having centroids within 0.05 pixels of a
detector pixel center. Next, we experimented with a variety of two-dimensional
functional forms to represent the PSF, progressively adding parameters until
finding one with a good match to the data. Since in our case the detector pixel
width is comparable to the PSF full-width half maximum value, it was necessary
to take into account not only the effect of the finite detector pixel area in
sampling the function, but also intra-pixel sensitivity variations.

Charge diffusion is the largest contribution to non-uniform sensitivity within
any given pixel. During the technology development phase of a space mission to
survey extragalactic supernovae,~\citet{brown} measured the effect of charge
diffusion on the intra-pixel sensitivity of a HAWAII-2 detector. He found a
good empirical fit to a typical pixel's response by convolving a tophat
function (with width equal to that of the detector pixel) with a hyperbolic
secant diffusion term, $\sech(\frac{r}{l_d})$, where r is radius from origin
and $l_d$ is the diffusion length. With the established diffusion length of 1.9
$\mu$m (compared to the 18 $\mu$m full pixel width), the response falls to
about 50\% of the peak at the middle of each pixel edge. For lack of similar
measurements of our own HAWAII-2 detector, we assumed the same charge diffusion
behavior.

We discretized the two-dimensional functions representing the PSF,
$\mathcal{P}_M(u, v)$, and intra-pixel response, $\mathcal{R}_M(i, j)$, at a
resolution of $M$ times that of the detector, where $M$ is an odd number $\ge$
3. In other words, the image model has $M \times M$ samples contained within
each detector pixel, one always aligned with the center of a pixel. The
intra-pixel response function, $\mathcal{R}_M$, is only defined over an area of
one pixel, so that $i, j \in \lbrace0, 1,\dotsc,M-1\rbrace$, whereas
$\mathcal{P}_M$ is defined over the entire $9M\times9M$ area of the mean PSF
cutout, corresponding to $u, v \in \lbrace0, 1, \dotsc, 9M-1\rbrace$. In this
notation, the detector-downsampled PSF, $\mathbb{P}(x, y)$ is determined by

\begin{equation} \mathbb{P}(x, y) = \sum_{i = 0}^{M-1} \sum_{j = 0}^{M-1} \mathcal{P}_M(i + Mx, j + My)\mathcal{R}_M(i, j)\label{eqn:psfbinning}\end{equation}

\noindent where $x, y \in \lbrace0, 1,\dotsc,8\rbrace$ are independent
variables representing detector samples over the $9\times9$ pixel mean PSF
cutout. 

We found a satisfactory functional form to match the monochromatic PSF by
taking the sum of two piecewise, two-dimensional Gaussian profiles, defined as
follows:

\begin{equation} \mathcal{P}_M(u,v) = \left\{
				 \begin{array}{lr}
					Ae^{-\left(\frac{r(u,v)^2}{2\sigma^2_{A,r}} + \frac{s(u,v)^2}{2\sigma^2_{A,s+}}\right)}
					+ Be^{-\left(\frac{r(u,v)^2}{2\sigma^2_{B,r}} + \frac{s(u,v)^2}{2\sigma^2_{B,s+}}\right)} & s > 0\\
					Ae^{-\left(\frac{r(u,v)^2}{2\sigma^2_{A,r}} + \frac{s(u,v)^2}{2\sigma^2_{A,s-}}\right)}
					+ Be^{-\left(\frac{r(u,v)^2}{2\sigma^2_{B,r}} + \frac{s(u,v)^2}{2\sigma^2_{B,s-}}\right)} & s \le 0
				\end{array}
				\right.
\label{eqn:psfdef}
\end{equation}

where \begin{equation}
\begin{bmatrix}
r\\
s
\end{bmatrix} = 
\begin{bmatrix}
\cos\theta & \sin\theta \\
-\sin\theta & \cos\theta
\end{bmatrix}
\begin{bmatrix}
\frac{1}{M}\left(u - \frac{9M-1}{2}\right)\\
\frac{1}{M}\left(v - \frac{9M-1}{2}\right)
\end{bmatrix}
\label{eqn:coordxform}
\end{equation}

\noindent Nine parameters fully describe the PSF in this formulation: two
amplitudes, six characteristic widths, and one rotation. The piecewise
definition allows freedom from reflective symmetry across the $r$ axis. Hence
there is a pair of characteristic widths for each side of the $s$ axis---one for
$s > 0$ ($\sigma_{A,s+}$ and $\sigma_{B,s+}$) the other for $s \le 0$
($\sigma_{A,s-}$ and $\sigma_{B,s-}$). Also note the coordinate transformation
built into the definition (Equation~\ref{eqn:coordxform}). The operations in
the right hand column vector serve two purposes. First, they shift the
effective origin from the lower left corner of the $9\times9$ cutout
array---its original position for the purpose of simplified indexing in
Equation~\ref{eqn:psfbinning}---to its center. At the same time, the
$\frac{1}{M}$ factor scales both coordinates to units of detector pixel width.
Finally, the $2\times2$ matrix facilitates a rotation of the overall surface by
angle $\theta$ in the counterclockwise sense.

\begin{figure}[h]
\epsscale{0.57}
\plotone{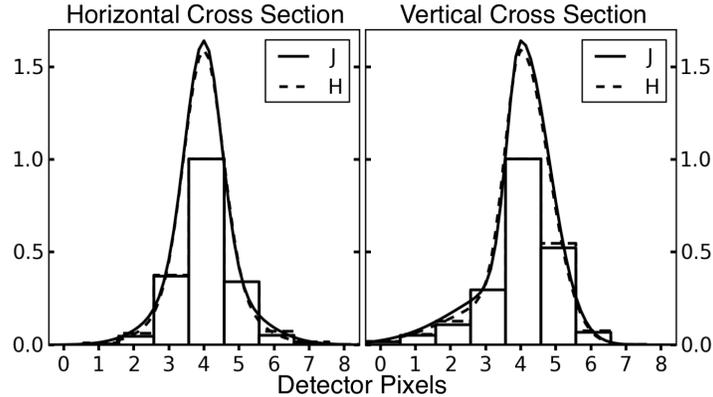}

\caption{Orthogonal cross sections through the center of the IFU PSF fit for
wavelengths 1.25 $\mu$m ($J$) and 1.58 $\mu$m ($H$). The curves illustrate the
function given in Equation~\ref{eqn:psfdef} with the parameters given in
Table~\ref{tab:psfparams}. The bars represent the same models after being
downsampled to the detector resolution using the assumed intra-pixel response.}

\label{fig:psf_cross_sections}
\end{figure}

Using MPFIT, the non-linear least squares fitting program written
by~\cite{markwardt}, we determined the function parameters for the mean PSF
cutouts at wavelengths 1.25 $\mu$m and 1.58 $\mu$m. The results, based on a
model spatial sampling rate of $M = 11$ times that of the detector, are listed
in Table~\ref{tab:psfparams}. In each case, the amplitudes were scaled so as to
give unity peak intensity in the detector-downsampled PSF. As in
Equation~\ref{eqn:psfdef}, the characteristic widths are in units of detector
pixel widths. In Figure~\ref{fig:psf_cross_sections} we have plotted orthogonal
cross sections of the best-fit PSF functions. In the same figure we drew bars
to represent the corresponding detector-downsampled PSF cross sections. Note
that the peak of each model function is significantly higher than that of the
detector-downsampled version, due to the sensitivity roll-off away from the
pixel center. At both wavelengths, the mean residual disparity between the
downsampled best-fit model and the original mean laser PSF cutout (not shown in
the plot) is the less than $1\%$ of the peak intensity. 

\begin{deluxetable}{cccccccccc}
\tabletypesize{\footnotesize}

\tablecaption{Parameters describing the P1640 IFU PSF at two wavelengths, as
defined in Equation~\ref{eqn:psfdef}.\label{tab:psfparams}}

\tablehead{\colhead{Wavelength ($\mu$m)} & \colhead{A} & \colhead{B} &
\colhead{$\sigma_{A,u}$} & \colhead{$\sigma_{B,u}$} & \colhead{$\sigma_{A,v+}$}
& \colhead{$\sigma_{A,v-}$} & \colhead{$\sigma_{B,v+}$} &
\colhead{$\sigma_{B,v-}$} & \colhead{$\theta$}}

\startdata

1.25 & 1.31 & 0.33 & 0.52 & 0.85 & 0.38 & 1.22 & 0.91 & 1.85 & $18.2^{\circ}$ \\
1.58 & 1.32 & 0.27 & 0.55 & 0.81 & 0.44 & 1.14 & 0.90 & 1.87 & $14.6^{\circ}$ \\

\enddata
\end{deluxetable}

\subsection{Spectrum Image Model}
\label{sec:spectrummodel}

We built upon knowledge of the spectrograph PSF to characterize the coarse
near-infrared spectra distributed across the IFU focal plane. Here we turned to
our Moon and twilight sky calibration exposures, during which each microlens
was illuminated with a strong, uniform, broad spectrum of light. An example
image of this kind was illustrated in Figure~\ref{fig:focpldiagram}. By
isolating the small detector area containing an individual microlens signal, we
can fit a set of parameters encoding the spectrum geometry. 

\begin{deluxetable}{ccccc}
\tabletypesize{\footnotesize}
\tablecaption{Definitions of parameters describing the geometry of an individual spectrum image. The mean, range, and standard deviation values of the parameters, as determined for the September 2009 focal plane, are also listed.\label{tab:specparams}}
\tablehead{\colhead{Parameter} & \colhead{Definition} & \colhead{Mean} & \colhead{Range} & \colhead{Std. Dev.}}
\startdata

Position $\left(X_0,Y_0\right)$ & $\left(X_{1.37 \mu\mbox{\tiny m}}, Y_{1.37 \mu\mbox{\tiny m}}\right)$ & N/A & (0.0--2047.0, 0.0--2047.0) & N/A \\
Height ($h$) & $Y_{1.10 \mu\mbox{\tiny m}} - Y_{1.76 \mu\mbox{\tiny m}}$ & 23.9 pixels & 23.5--24.7 pixels & 0.2 pixels \\
Tilt ($t$) & $\left(X_{1.10 \mu\mbox{\tiny m}} - X_{1.76 \mu\mbox{\tiny m}}\right)/\left(Y_{1.10 \mu\mbox{\tiny m}} - Y_{1.76 \mu\mbox{\tiny m}}\right)$ & 0.044 & 0.0090--0.080 & 0.021 \\

%Amplitude ($a$) & Mean-normalized amplitude & 1.00 & 0.29--1.49 & 0.27 \\

\enddata
\end{deluxetable}

In Table~\ref{tab:specparams} we have listed the parameters needed to describe
the spectrum image of an individual microlens. The $\left(X_0,Y_0\right)$
position coordinates are the most fundamental of these. They are referenced to
the $\lambda$ = 1.37 $\mu$m point, which coincides with the sharp (blue) edge
of the telluric water absorption trough between the $J$ and $H$ bands. The $X,
Y$ coordinates index the full detector array from an origin at the pixel in the
lower left (SW) corner of the image; all integral values align with a pixel
center. Using similar notation, we defined the spectrum height and tilt based
on the relative positions of the $\lambda$ = 1.10 $\mu$m and 1.76 $\mu$m
points, which roughly correspond to the edges of our passband. From the
position, height, and tilt, the coordinates of an arbitrary wavelength in the
spectrum can be calculated by the following parametrized equations:

\begin{equation}
\begin{array}{lr}
X(w) = X_0 + th\left(\frac{9-w}{22}\right)\\
Y(w) = Y_0 + h\left(\frac{9-w}{22}\right)
\end{array}
\label{eqn:specxyparam}
\end{equation}

\noindent where $h$ is height, $t$ is tilt, and $w = \left(\lambda -
1.1\:\mu\mbox{m}\right)/0.03\:\mu$m. By this definition, each integral step in
$w$ corresponds to 0.03 $\mu$m, so there is a total of 22 such increments from
1.10 $\mu$m and 1.76 $\mu$m. Using the same wavelength parameter, we can
specify the intrinsic spectrum incident on the focal plane by some function
$s(w)$ for $0 \le w \le 22$.

%The last parameter listed in Table~\ref{tab:specparams},
%amplitude ($a$) is based on the overall signal strength associated with a given
%microlens. It is determined during the fitting procedure described later on in
%this section. 

\begin{figure}[tb]
\epsscale{0.9}
\plotone{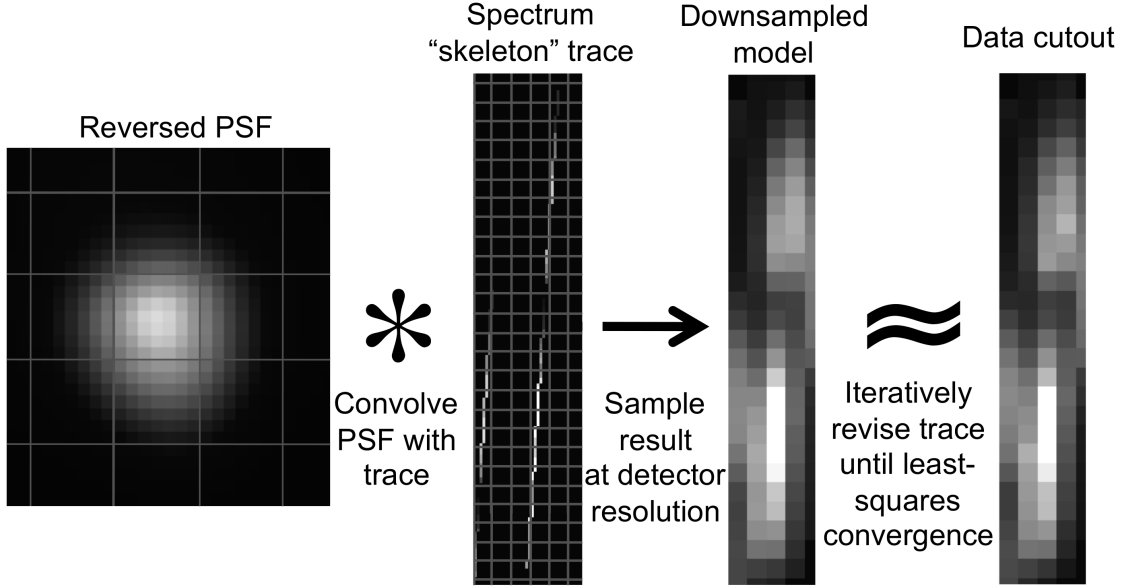}

\caption{Diagram summarizing the microlens spectrum modeling procedure. On the
far left, the IFU PSF model (reflected about the center) is shown with a
spatial sampling factor 7 times that of the detector, with an overlaid grid of
gray lines representing pixel boundaries. We convolve this with the spectrum
model trace shown in the next panel (again shown with a gray grid to illustrate
the scale of detector pixels) in order to simulate the true light distribution.
To test the accuracy of the trace, we downsample the convolution result to the
detector resolution, and compare it to the data cutout being fit. The fitting
algorithm switches between adjusting the trace parameters and repeating the
convolution and downsampling procedure until the model converges.}

\label{fig:moddescrip}
\end{figure}

We introduce the concept of spectrum \emph{trace} to model the layout of the
spectrum by an ideal ``skeleton'' image formed by a train of impulse functions,
unencumbered by diffraction and focus effects. The trace is discretized in the
same manner as the PSF model, at a resolution $M$ times higher than that of the
detector. In this case, however, we use a lower spatial sampling rate factor of
$M = 7$ to balance reasonable execution speed and performance. By convolving
the trace with the reverse of the PSF (rotated 180$^\circ$), and downsampling
the result, we can synthesize a spectrum image as measured by the detector
(Figure~\ref{fig:moddescrip}). The position, height, and tilt parameters, along
with the intrinsic spectrum can then be adjusted by a least squares fitting
algorithm until the downsampled result matches the data cutout. 

A cutout spanning an area of $9\times29$ detector pixels is sufficient to
enclose an individual spectrum as well as major portions of the two nearest
neighbors. At the start of the fitting procedure, the cutout is aligned such
that the ($X_0, Y_0$) reference point of the spectrum nearly corresponds to
pixel position (4, 16) in the detector-downsampled trace array. Two free
parameters in the model, $\Delta X$ and $\Delta Y$, allow the algorithm to
refine the initial position guess alongside the other geometrical properties.
For a given microlens spectrum, the following equations define the conversion
between $X, Y$ detector indices and $u, v$ trace array indices:

\begin{equation}
\begin{array}{lr}
u = \left( X - X_0 + \frac{9}{2} + \Delta X\right) M \\
v = \left( Y - Y_0 + \frac{33}{2} + \Delta Y\right) M \\
\end{array}
\label{eqn:uvdef}
\end{equation}

\noindent from which it follows that the trace indices of the 1.37 $\mu$m
reference point are $\left(u_0, v_0\right) = \\ \left( M \left(\frac{9}{2} +
\Delta X\right), M \left(\frac{33}{2} + \Delta Y\right) \right)$. Based on the
typical interval between spectra along a microlens, we set the 1.37 $\mu$m
reference points of the neighboring spectra by $(u_{0\uparrow}, v_{0\uparrow})
= (u_0 + 3.3M, v_0 + 10.0M)$ and $(u_{0\downarrow}, v_{0\downarrow}) = (u_0 -
3.3M, v_0 - 10.0M)$.

The trace signal is dispersed over the same line segment defined in Equation
set~\ref{eqn:specxyparam}, with intrinsic spectrum function $s(w)$. The
neighboring spectra are parametrized by the same shape with respect to their
own reference points, ($u_{0\uparrow}, v_{0\uparrow}$) and ($u_{0\downarrow},
v_{0\downarrow}$). We form separate trace arrays for the $J$- and $H$-band
halves of the spectra, designated $\mathcal{T}_{J,M}(u,v)$ and
$\mathcal{T}_{H,M}(u,v)$. We do this in anticipation of separate convolution
operations with the $J$- and $H$-band PSFs
(Figure~\ref{fig:psf_cross_sections}). The two respective trace arrays are
defined as follows:

\begin{equation}
\begin{array}{lr}

\mathcal{T}_{J,M}(u,v) = \left\{
\begin{array}{ll}
s(w) & -\frac{1}{2} < u - u_0 - t(v - v_0) \le \frac{1}{2} \;\;\;\mbox{and}\;\;\; 0 \le w < 10 \\
s(w_\uparrow) & -\frac{1}{2} < u - u_{0\uparrow} - t(v - v_{0\uparrow}) \le \frac{1}{2} \;\;\;\mbox{and}\;\;\; 0 \le w_\uparrow < 10 \\
s(w_\downarrow) & -\frac{1}{2} < u - u_{0\downarrow} - t(v - v_{0\downarrow}) \le \frac{1}{2} \;\;\;\mbox{and}\;\;\; 0 \le w_\downarrow < 10 \\
0 & \mbox{otherwise} \\
\end{array}
\right. \\

\mathcal{T}_{H,M}(u,v) = \left\{
\begin{array}{ll}
s(w) & -\frac{1}{2} < u - u_0 - t(v - v_0) \le \frac{1}{2} \;\;\;\mbox{and}\;\;\; 10 \le w \le 22 \\
s(w_\uparrow) & -\frac{1}{2} < u - u_{0\uparrow} - t(v - v_{0\uparrow}) \le \frac{1}{2} \;\;\;\mbox{and}\;\;\; 10 \le w_\uparrow \le 22 \\
s(w_\downarrow) & -\frac{1}{2} < u - u_{0\downarrow} - t(v - v_{0\downarrow}) \le \frac{1}{2} \;\;\;\mbox{and}\;\;\; 10 \le w_\downarrow \le 22 \\
0 & \mbox{otherwise}
\end{array}
\right. \\

\end{array}
\end{equation}

\noindent where $w = 9 - \frac{22}{Mh}\left(v - v_0\right)$, $w_{\uparrow} = 9 - \frac{22}{Mh}\left(v - v_{0\uparrow}\right)$, and $w_{\downarrow} = 9 - \frac{22}{Mh}\left(v - v_{0\downarrow}\right)$. 

To test a given set of spectrum model parameters, we convolve the $J$ and $H$
trace arrays with the reverse of the PSF model (such that
$\mathcal{P}^\prime_M(i,j) = \mathcal{P}_M(-i,-j)$), giving a high-resolution
model of the light distribution on the focal plane, $\mathcal{S}_M(u, v)$:

\begin{equation}
\mathcal{S}_M = \mathcal{T}_{J,M} * \mathcal{P}^\prime_{J,M} + \mathcal{T}_{H,M} * \mathcal{P}^\prime_{H,M}
\label{eqn:psfconv}
\end{equation}

\noindent Implicitly, we have zero-padded the trace arrays before the
convolution, and trimmed the result to the original $9M\times29M$ array size.
The detector-downsampling operation is similar to that used earlier for the PSF
model:

\begin{equation}
\mathbb{S}(x,y) = b + \sum_{i = 0}^{M-1} \sum_{j=0}^{M-1}\mathcal{S}_M(i + Mx, j + My)\mathcal{R}_M(i,j) 
\label{eqn:binnedspec}
\end{equation}

\noindent However, one new variable has been introduced in
Equation~\ref{eqn:binnedspec}: $b$, a constant offset added to each pixel in
the downsampled image model.  This is one more parameter open to adjustment by
the fitting procedure, which takes into account any background level of
scattered light present in the data cutout. For a point source, in some focal
plane locations this background reaches up to 3\% of the 99.5 percentile-level
count rate, considering all detector pixels. Therefore, it becomes especially
significant for an exposure of a source as bright as the Moon. The resulting
synthetic spectrum image $\mathbb{S}$ can be directly compared with the data
cutout (Figure~\ref{fig:moddescrip}). In principle, the least squares fitting
algorithm (the MPFIT program in our case) converges on the data cutout over
many loops, switching between revising the trace model parameters and comparing
the downsampled result to the data.

The combination of unknown position, shape, and intrinsic spectrum $s(w)$
presents too many free parameters for a fitting algorithm to accurately solve
for at once. In practice, we need to iteratively build up constraints, starting
from as few assumptions as possible. One aspect of the Moon/sky calibration
exposure we can take advantage of is the fact that the intrinsic spectrum,
$s(w)$, is identical across the image, apart from scale factors due to
large-scale variations in sensitivity over the field of view. In addition, by
referring to the laser calibration images, we can make very good initial
guesses of the height and tilt for a given region of the focal plane.  Still,
we found these constraints alone were insufficient to reach consistent
solutions. The exact vertical position of the spectrum (encoded by $\Delta Y$
in Equation Set~\ref{eqn:uvdef}) proved especially difficult to determine with
only limited information about the light source and the instrument response. To
get over this barrier, we chose to use prior knowledge of the atmosphere's
transmission function---in particular, the shape imposed on the spectrum by the
deep water absorption trough in the middle of the P1640 passband.

\begin{figure}[tb]
\epsscale{0.4}
\plotone{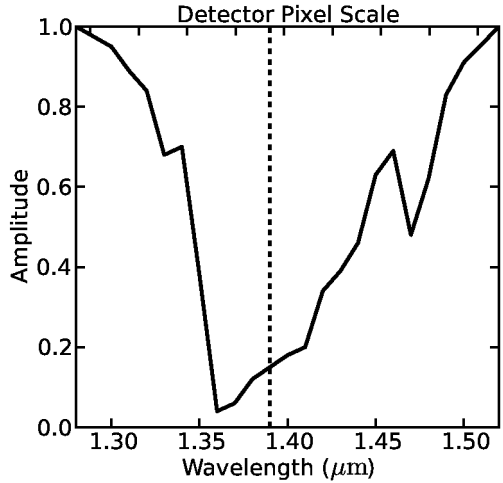}

\caption{The atmospheric transmission function around the water absorption
trough between the $J$ and $H$ bands, binned to 0.01 $\mu$m resolution, based
on the measurements by~\cite{manduca}. The tick marks on the top axis indicate
the scale of vertical detector pixels along the dispersion axis of the
spectrum. During the spectrum fitting procedure, when the exact position and
shape have not yet been established, the model trace spectrum $s(w)$ is forced
to follow this profile in the range $\lambda = \mbox{1.30--1.50}\:\mu\mbox{m}$.
The dotted line separates those points of $s(w)$ that are fixed relative to the
free spectrum value at $\lambda = 1.28\:\mu$m ($6 < w < 10$) versus $\lambda =
1.52\:\mu$m ($10 \le w < 14$).}

\label{fig:watertrough}
\end{figure}

Figure~\ref{fig:watertrough} shows the expected transmission function of the
atmosphere from 1.28 $\mu$m to 1.52 $\mu$m. The data points are based on the
measurements made by~\cite{manduca} from Kitt Peak (at altitude 6875 ft,
comparable to the 5618 ft altitude of the Palomar Observatory Hale Telescope),
here averaged over 0.01 $\mu$m bins. Instead of allowing the points inside the
water trough ($6 < w < 14$) to vary freely, we impose the condition

\begin{equation}
s(w) = \left\{
\begin{array}{ll}
T_{\mbox{\tiny atm}}(w)s(6) & 6 < w < 10 \\
T_{\mbox{\tiny atm}}(w)s(14) & 10 \le w < 14 \\
\end{array}
\right.
\label{eqn:troughcond}
\end{equation}

\noindent where $T_{\mbox{\tiny atm}}(w)$ is the peak-normalized atmospheric
transmission function plotted in Figure~\ref{fig:watertrough}. Inside the water
trough, $s(w)$ is discretized in 0.01 $\mu$m bins; outside, in 0.03 $\mu$m bins
(integral values of $w$). Once the above assertion is in place, the fitting
algorithm could at last reliably determine both the position and shape of the
spectrum, as achieved in the example shown in Figure~\ref{fig:moddescrip}.

\label{eqn:watercond}

\subsection{Global Spectrograph Focal Plane Solution}
\label{sec:globalsol}

We repeated the spectrum fitting procedure across the entire spectrograph image
to form a global solution unique to the specific epoch of the Moon/sky
calibration exposure. To minimize the number of free parameters before
executing this, we first determined a mean intrinsic spectrum $\bar{s}(w)$
based on average of the $s(w)$ fit results from a sub-area of about 100 spectra
near the center of the field of view. With the spectrum shape fixed, however,
there needs to be a parameter that captures variations in overall signal
strength across the focal plane. We designated an amplitude parameter $a$ to
acts as a multiplicative constant, applied to $\bar{s}(w)$, and freely adjusted
alongside $X_0$, $Y_0$, $h$, $t$, and $d$.

\begin{figure}
\epsscale{1.0}
\plotone{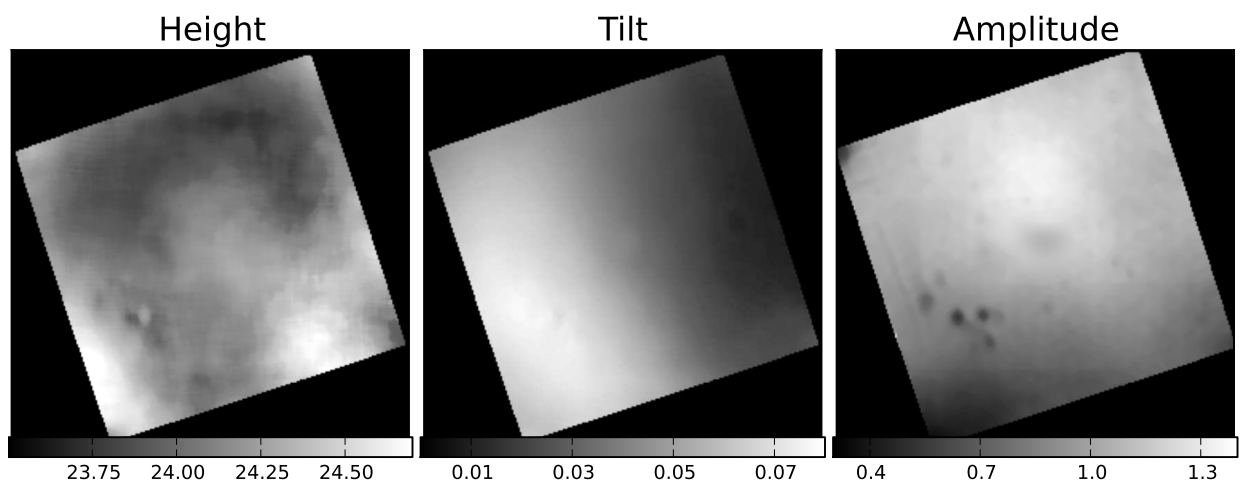}

\caption{Maps of spectrum parameters making up one global spectrograph focal
plane solution. This particular solution is based on the fit to the Moon
calibration image acquired on 28 June 2009. The maps appear rotated due to the
orientation of the microlens array with respect to the detector (see
Figure~\ref{fig:focpldiagram}). Height is shown in units of detector pixels,
tilt is an inverse slope (unitless), and amplitude is mean-normalized.}

\label{fig:param_maps}
\end{figure}

In Figure~\ref{fig:param_maps} we have displayed maps of the height, tilt, and
amplitude parameters for one epoch. These maps proved essential to the
challenging process of debugging the fitting routines. They also enable easy
visual comparisons between focal plane properties at different times, and can
serve as diagnostic tools during periods of modifications and upgrades to
instrument optics. The maps in Figure~\ref{fig:param_maps} appear as rotated
squares because the microlens array is by design rotated with respect to the
detector (as shown previously in Figure~\ref{fig:focpldiagram}). We index the
microlenses using Cartesian coordinates $i$ and $j$ relative to an origin at
the lower left corner. With these coordinates, a range of $0 \le i,j < 250$ is
sufficient to enclose the $3.8\times10^4$ microlens spectra on the detector.

\begin{figure}
\epsscale{1.0}
\plotone{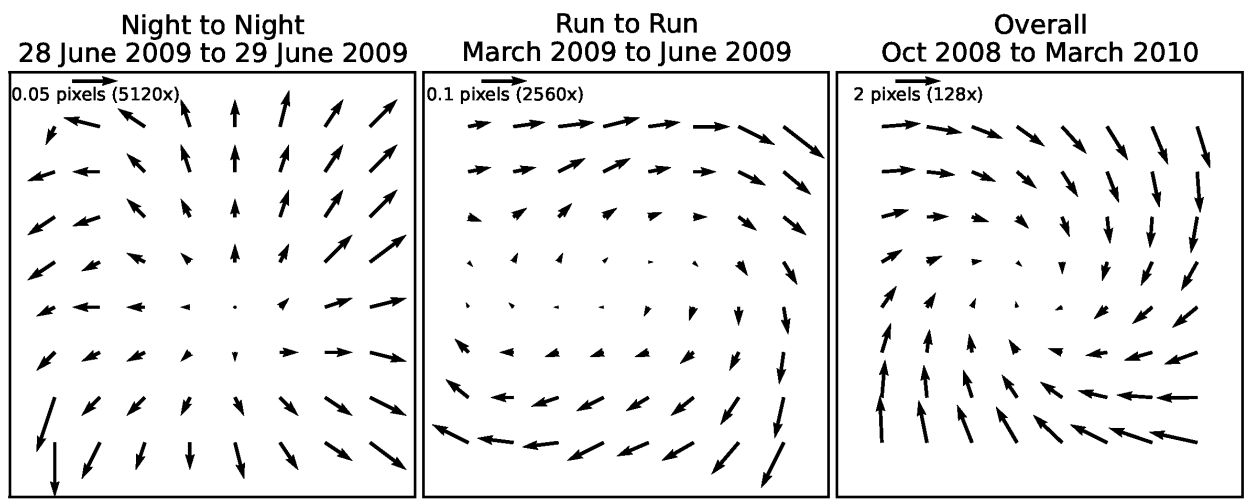}

\caption{Vector field plots illustrating examples of the evolution of the
spectrograph focal plane geometry over three time intervals: 1 day, 3 months,
and 17 months.  Each quiver represents the median change in spectrum position
inside a box $1/8^{\mbox{\footnotesize th}}$ of the full image width. In the
top left corner of each plot, a small legend indicates the relative scale of
the quiver in terms of detector pixel widths.  For each comparison, we
customized the quiver scale factor to clearly reveal the transformation.}

\label{fig:distortion}
\end{figure}

The ability to analyze the \emph{spatial} distribution of the IFU spectra is
also of great interest. A vector field, like those depicted in
Figure~\ref{fig:distortion}, is an effective way to illustrate the evolution of
the global spectrograph focal plane geometry. To make these plots, we first
partitioned the calibration image into an array of $8\times8$ boxes, each of
width 256 detector pixels. For any two comparison epochs, each with spectrum
position arrays $X_0(i,j)$ and $Y_0(i,j)$, we calculated the median of the
differences $\Delta X_0(i,j)$ and $\Delta Y_0(i,j)$ inside each box, resulting
in an $8\times8$ array of vectors. Before plotting those vectors, we subtracted
the median difference vector at the image center. In this way, we removed the
effect of a trivial bulk shift between the focal plane patterns.

Depending on the duration of time between the pair of calibration images under
consideration, the quivers representing the vectors need to be scaled up by
different factors to reveal the subtle evolution. Once this is done, it becomes
clear that the overall scale and orientation of the spectrograph focal plane
pattern vary with time. More complicated, non-uniform distortions also play a
role. In all cases, the magnitude of these changes are small enough that they
would never be obvious from a mere ``blinking'' comparison of the source
images. For example, the transformation from March 2009 to June 2009 can mostly
be attributed to a rotation of the microlens array with respect to the detector
(or vice versa) by an angle of merely $14''$. Likewise, between 28 June 2009
and 29 June 2009, the focal plane was magnified by about 0.003\%. It is a fair
guess that the changes we observe in global focal plane geometry are due to
minute variations of environmental conditions inside the IFU dewar.  However,
it remains unclear what the relative contributions are from the various optics
and the mechanical support structures. To an extent, the origin of these
changes is not an issue, so long as each science data set can be attached to a
solution that accurately reflects its particular geometry.

Another important product of our global fitting procedure is a synthetic image
of the entire spectrograph focal plane. Using the established geometric
parameters, we can inject an arbitrary source spectrum $s_\star(w)$ to simulate
the distribution of light incident on the detector. The synthetic focal plane
image is useful for inspecting the results of the global fit, and is also an
essential ingredient in the algorithm used by the cube extraction pipeline to
register the spectrograph image of an individual science exposure at sub-pixel
precision (described in \S~\ref{sec:specreg}). The formalism is analogous to
that described for the individual spectrum cutout model in
Equations~\ref{eqn:uvdef}--\ref{eqn:binnedspec}. We designate $\mathcal{F}_J(p,
q)$ and $\mathcal{F}_H(p, q)$ to represent the $J$- and $H$-band model trace
arrays of the full spectrograph focal plane image, discretized at a spatial
sampling rate $M$ times that of the detector. Since the HAWAII-2 detector array
size is $2048\times2048$, the trace arrays are defined over $p, q \in \lbrace0,
1, \dotsc, 2048M-1\rbrace$. Here we again settled on a sampling factor of $M$ =
7. Since as before, we require $M$ to be an odd integer $\ge3$, there is always
a pair of trace array indices $p,q$ aligned with the center of a given detector
pixel $X, Y$:

\begin{equation}
\begin{array}{lr}
p = MX + \frac{M-1}{2}\\
q = MY + \frac{M-1}{2}.\\
\end{array}
\end{equation}

\noindent The trace array is determined by the position, height, and tilt
solutions, now indexed by microlens coordinates $i,j$:

\begin{equation}
\begin{array}{lr}

\mathcal{F}_{J,M}(p,q) = {\displaystyle\sum\limits_{i = 0}^{249} \sum\limits_{j=0}^{249} \mathcal{S}_{J,M,i,j}(p,q)}\\

\mathcal{F}_{H,M}(p,q) = {\displaystyle\sum\limits_{i = 0}^{249} \sum\limits_{j=0}^{249} \mathcal{S}_{H,M,i,j}(p,q)}\\

\end{array}
\end{equation}

\begin{equation}
\begin{array}{l}
\mbox{where} \\

\begin{array}{lr}
\mathcal{S}_{J,M,i,j}(p,q) = \left\{
\begin{array}{ll}
s_\star(w_{ij}) & -\frac{1}{2} < p - p_{0_{ij}} - t_{ij}(q - q_{0_{ij}}) \le \frac{1}{2} \;\;\;\mbox{and}\;\;\; 0 \le w_{ij} < 10 \\
0 & \mbox{otherwise} \\
\end{array}
\right. \\

\mathcal{S}_{H,M,i,j}(u,v) = \left\{
\begin{array}{ll}
s_\star(w_{ij}) & -\frac{1}{2} < p - p_{0_{ij}} - t_{ij}(q - q_{0_{ij}}) \le \frac{1}{2} \;\;\;\mbox{and}\;\;\; 10 \le w_{ij} \le 22 \\
0 & \mbox{otherwise}
\end{array}
\right. \\
\end{array}\\

p_{0_{ij}} = MX_{0_{ij}} + \frac{M-1}{2},\quad q_{0_{ij}} = MY_{0_{ij}} + \frac{M-1}{2},\quad\mbox{and}\:w_{ij} = 9 - \frac{22}{Mh_{ij}}\left(q - q_{0_{ij}}\right).\\

\end{array}
\end{equation}

From the trace arrays, we obtain the spectrograph focal plane image model
$\mathcal{I}_M(p,q)$ in the same manner as in \S~\ref{sec:spectrummodel}, by
convolving them with their corresponding reversed PSF models:

\begin{equation}
\mathcal{I}_M = \mathcal{F}_{J,M} * \mathcal{P}^\prime_{J,M} + \mathcal{F}_{H,M} * \mathcal{P}^\prime_{H,M}
\label{eqn:imagemodel}
\end{equation}

\noindent We implement these convolution operations in the Fourier domain to
save computational time, which is otherwise a nuisance for the large dimensions
of our arrays~\citep[chap.  6]{bracewell}. With a spatial sampling factor of $M
= 7$, we obtain a factor of $\sim$ 50 using FFT-based convolution over direct
convolution. In our experience, this cuts the execution time needed to form the
synthetic image down from a few hours to a few minutes (assuming the global
solution is already done).

Finally, we can obtain the detector-downsampled synthetic focal plane image,
$\mathbb{I}(X,Y)$, by binning $\mathcal{I}_M$ to the detector resolution using
the assumed intra-pixel response:

\begin{equation}
\mathbb{I}(X,Y) = \sum_{i = 0}^{M-1} \sum_{j=0}^{M-1}\mathcal{I}_M(i + MX, j + MY)\mathcal{R}_M(i,j) 
\label{eqn:binnedimagemodel}
\end{equation}

\noindent Note that this synthetic detector image is idealized in the sense
that we have left out the amplitude modulations across the image ($a_{ij}$) as
well as background light parameters ($d_{ij}$). For the purpose of registering
a science spectrograph image, this is preferred, since we are only concerned
with matching the shapes and positions of the spectra.

\section{Data Cube Extraction Pipeline}

%\subsection{Overview}

\begin{figure}[h]
\epsscale{0.4}
\plotone{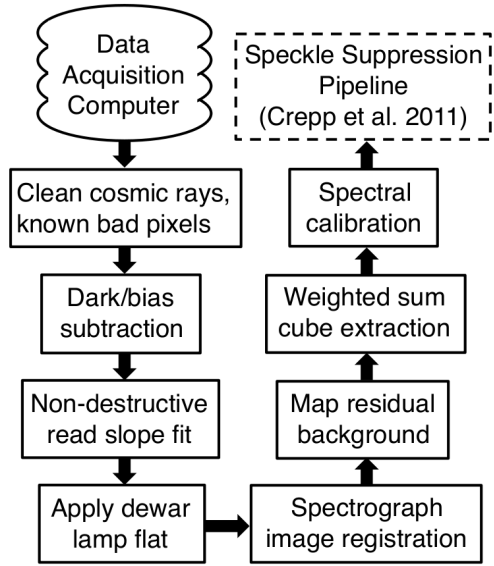}

\caption{Block diagram of the Project 1640 Data Cube Extraction Pipeline
(PCXP). Raw detector data are processed into reduced, registered spectrograph
images (\S~\ref{sec:detectorproc}). In the second main loop, beginning with the
residual background mapping, the spectrograph images are translated to data
cubes with the help of the global spectrograph focal plane solution
(\S~\ref{sec:cubeextraction}).}

\label{fig:blockdiagram}
\end{figure}

The Project 1640 Data Cube Extraction Pipeline (PCXP), written in the GNU C
programming language, automates the processing of raw P1640 detector images and
their translation to reduced data cubes. A block diagram summarizing the steps
applied to each image is shown in Figure~\ref{fig:blockdiagram}. By design, the
program is fast enough to use while observing, so that newly acquired images
can be inspected in real time to monitor instrument performance and check for
unknown objects. The data pipeline can also be used to process an arbitrarily
large set of raw data at a later date. For post-processing, we feed the output
of the PCXP into the Project 1640 Speckle Suppression Pipeline (PSSP),
described by~\cite{creppSSP}. 

Two outer loops comprise the PCXP execution. First, the program steps through
the detector data in the input directory specified by the user at the start
time, processing each non-destructive read (NDR) sequence to form a reduced,
registered spectrograph image. The second stage of the pipeline loops through
the finished spectrograph images and extracts data cubes from each of them.
Throughout these steps, the pipeline relies on the empirical model of the
spectrograph focal plane described in \S~\ref{sec:focplmodel}. 

\subsection{Detector Image Processing}

\label{sec:detectorproc}

\subsubsection{Cosmic Ray Removal}

The pipeline identifies pixels contaminated by cosmic rays by checking for
anomalous jumps in digitized count values within the NDR sequence. For each
detector pixel, our algorithm determines the median increase in counts between
successive reads over the course of the exposure. A count increment greater
than five times the median is flagged as a cosmic ray event. At each pixel
meeting this criterion, the count contribution of the cosmic ray event is
subtracted from the read corresponding to the event as well as all the
following reads, canceling out its influence. We chose our threshold based on
inspections of images of faint occulted stars, with relatively noisy slopes. We
blinked ``before'' and ''after'' images to check that all apparent cosmic ray
events, and no starlight-dominated pixels were erroneously flagged. 

This method of cosmic ray removal only works for exposures consisting of more
than two reads. For a shorter exposure there is no way to take advantage of the
NDR detector mode to identify cosmic rays. In this case the pipeline passes the
detector image through the IRAF NOAO cosmic ray cleaning algorithm.

\subsubsection{Bias/Dark Subtraction}

\begin{figure}[h]
\epsscale{0.4}
\plotone{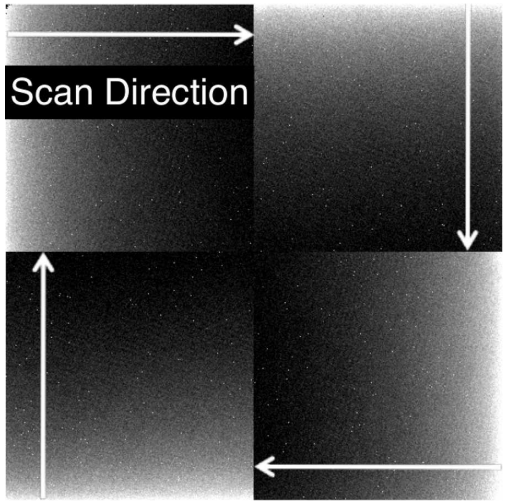}

\caption{An example of a dark exposure used to subtract the bias, thermal
counts, and hot pixels from science images. The ``bias tilt''---the gradient in
the bias pedestal---is strongest along the scan direction of each detector
quadrant (each with its own read-out amplifier).} 

\label{fig:dark}
\end{figure}

During each observing run, a set of ``dark'' NDR sequences are obtained by
taking calibration exposures with the IFU in a cryogenic state identical to the
scientific data acquisition mode, except that the coronagraph beam entrance
window is capped to obstruct external light. These dark exposures record the
bias, thermal, dark current, and badly-behaved ``hot'' pixel count values of
the detector array at each read interval. The median of 11 dark NDR sequences
for each exposure time is added to a permanent library directory of dark
exposures, marked by date and exposure time. See Figure~\ref{fig:dark} for an
example of a dark exposure. After loading the \textit{dat} file of a science
target, the first processing step of the pipeline is to find the most
appropriate dark NDR sequence and perform a readwise subtraction. 

\subsubsection{Non-destructive Read Slope Fitting}
\label{sec:NDRslope}

After subtracting the bias/dark component and removing the cosmic rays, the
pipeline fits a slope to the ADU count versus time values recorded in the NDR
sequence. This reduces the detector data for a given exposure to a single
$2048\times2048$ pixel representation of the spectrograph image. We employ an
ordinary least-squares linear regression to determine the count rate for each
pixel, eventually storing the floating point values in a FITS file in units of
counts/second.

The slope fitting is complicated by pixel saturation caused by bright sources.
However, the NDR detector mode is advantageous for handling this. In the case
where a pixel reaches saturation at some point after the first two reads, the
affected reads are simply excluded from that pixel's linear regression. This
approach, recommended after tests described by~\cite{ives}, extends the
effective dynamic range of a long exposure by a factor of $\sim N_{reads}/2$.
For pixels that reach saturation before the second read, a slope computation is
not possible. If this saturation occurred between the first and second reads,
then the slope can at least be approximated based on the difference between the
first read value and an assumed zero level from the dark NDR sequence. However,
in the case of a pixels saturating before the first read, this result will not
be physically meaningful. To prevent erroneous measurements from being made by
investigators analyzing images affected by saturation, the pipeline sets an
appropriate header variable in the reduced FITS files. This header keyword
indicates whether any of the detector pixels saturated, and if so, whether that
occurred at the first, the second, or a subsequent read.

\subsubsection{Detector Flat-Fielding}
\label{sec:dewarflat}

A externally controlled lamp inside the dewar of the P1640 IFU can fully
illuminate the detector. To counteract pixel-to-pixel variations in detector
sensitivity, we constructed a detector flat field map based on the mean of 12
dewar lamp exposures. Since the lamp intensity is not uniform across the
detector, we used IRAF to fit a cubic spline surface to the normalized, mean
dewar lamp image. We divided by the resulting spline surface to form the final
detector flat, with large-scale variations removed (see the next section for an
explantation of how large-scale variations in sensitivity are corrected for
during cube extraction). The standard deviation of pixel values in the detector
flat field map is 0.13. After the NDR slope-fitting step, the pipeline divides
the spectrograph image by the flat field map to compensate for pixel-to-pixel
variations. For locations in the flat field map with exceptionally low values
($< 0.3$), no division is carried out since doing so would tend to enhance the
noise induced by weak, problematic pixels.

\subsubsection{Spectrograph Image Registration}
\label{sec:specreg}

\begin{figure}[htb]
\epsscale{0.4}
\plotone{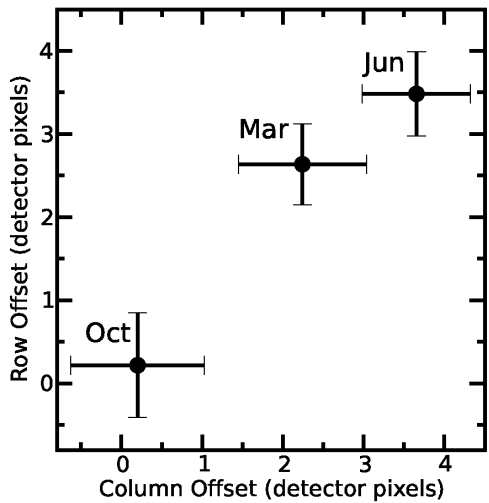}

\caption{Example distributions of the spectrograph-detector alignment offsets
with respect to a canonical template. The mean offsets, with bars indicating
the standard deviations in each direction, are plotted for three P1640
observing runs: October 2008, March 2009, and June 2009.}

\label{fig:offsets}
\end{figure}

Due to flexure---varying mechanical stress on the instrument while the
telescope slews---the projection of the microlens array onto the detector
changes over the course of an observing period. The plot in
Figure~\ref{fig:offsets} illustrates the magnitude of this effect based on
measurements from three observing runs. Between targets, the positions of the
spectra on the detector can uniformly shift by up to 2 pixels in each
direction. Between observing runs there is a more pronounced, systematic shift
in the spectrograph-detector alignment.  To accurately extract the data, the
pipeline needs to register the precise offset between each spectrograph image
and the focal plane model of the corresponding epoch. We accomplish this
through two stages: first a crude estimate based on a cross-correlation with
the downsampled spectrograph image model (array $\mathbb{I}$ in
Equation~\ref{eqn:binnedimagemodel}), followed by a more elaborate approach to
refine the offset to sub-pixel precision.

For efficiency, the initial cross-correlation is restricted to a $200 \times
200$ square pixel section of the science image $\mathbb{D}(X,Y)$. We denote
this cutout box with a tilde accent on top of the original array symbol:

\begin{equation}
\tilde{\mathbb{D}}(x,y) = \mathbb{D}\left(X_p - 100 + x, Y_p - 100 + y\right)\quad\mbox{for}\;0 \le x,y < 200.
\end{equation}

\noindent Likewise we use $\tilde{\mathbb{I}}$ to represent the same section
from the downsampled focal plane image model. The center of the box, $(X_p,
Y_p)$, is chosen based on the average count rate computed within $16 \times 16$
partitions across the image, so as to enclose spectra with relatively high
signal strength. In a typical science image with the star occulted by the
coronagraph, this is near the center of the image, where the residual starlight
is brightest.

We cross-correlate $\tilde{\mathbb{I}}$ and $\tilde{\mathbb{D}}$ to determine
the \emph{crude} offset. A given science focal plane pattern is not expected to
stray more than two pixels away from the calibration exposure of the matching
epoch.  Furthermore, the periodicity of the spectrum pattern ensures that large
lags will merely introduce degenerate solutions. Therefore, we do not compute
the full two-dimensional cross-correlation array, but merely a small region
bounded by horizontal and vertical lags up to 4 pixels in each direction:

\begin{equation}
\tilde{\mathbb{C}}_0(r,s) = {\displaystyle\sum\limits_{x=4}^{195}\sum\limits_{y=4}^{195}}\tilde{\mathbb{I}}(x - r, y - s)\tilde{\mathbb{D}}(x, y)\quad\mbox{for}\: -4 \le r,s \le 4
\end{equation}

\noindent The summation limits take the lag range into account in order to
avoid the influence of non-overlapping array edges. The lag combination that
maximizes $\tilde{\mathbb{C}}(r,s)$, which we denote $(r_p, s_p)$, is our
initial guess for the horizontal and vertical displacement of $\mathbb{D}$ with
respect to $\mathbb{I}$.

The second stage of the registration routine determines separately the
\emph{fine} $X$ and $Y$ offsets. As apparent in Figure~\ref{fig:focpldiagram},
the spectral dispersion is almost completely aligned with the $Y$ axis of our
detector image coordinate system. As a consequence, the shape of a spectrum's
horizontal cross-section at a given wavelength is determined much more by the
spectrograph PSF shape than by the intrinsic spectrum of the light source. It
is effectively the cross-dispersion profile commonly referred to in literature
on more conventional spectroscopic
observations~\citep[e.g,][]{crossdispersion}. Therefore, to measure the effect
of a slight horizontal offset on the detector-sampled image, we can start with
the high-resolution model of the light distribution $\mathcal{I}_M$
(Equation~\ref{eqn:imagemodel} from \S~\ref{sec:globalsol}), even though its
intrinsic spectrum does not necessarily match the data. To simulate how the
detector would ``see'' the image model for a range of small fractional-pixel
offsets from the initial alignment, we downsample $\mathcal{I}_M$ as in
Equation~\ref{eqn:binnedimagemodel}, but with the detector sampling array
shifted by a range of horizontal sub-pixel offsets indexed by the variable
integer $\delta$: 

\begin{equation}
\begin{array}{lr}
\tilde{\mathbb{I}}_\delta(x,y) = {\displaystyle\sum\limits_{i = 0}^{M-1} \sum\limits_{j=0}^{M-1}\mathcal{I}_M(M(X_p - 100 + x) + i - \delta, M(Y_p - 100 + y) + j)\mathcal{R}_M(i,j)}\\
{\displaystyle\mbox{for}\quad 0 \le x,y < 200\quad\mbox{and}\;-\frac{M-1}{2} \le \delta \le \frac{M-1}{2}.}\\
\end{array}
\end{equation}

\noindent Now we reevaluate the cross-correlation peak for the $M$ fine
horizontal offsets values, with $r$ and $s$ fixed at $r_p$ and $s_p$:

\begin{equation}
\tilde{\mathbb{C}}_\delta(r_p, r_s) = {\displaystyle\sum\limits_{x=4}^{195}\sum\limits_{y=4}^{195}}\tilde{\mathbb{I}}_\delta(x - r_p, y - s_p)\tilde{\mathbb{D}}(x, y)\quad\mbox{for}\;-\frac{M-1}{2} \le \delta \le \frac{M-1}{2}.
\end{equation}

\noindent The fractional offset index $\delta$ that maximizes
$\tilde{\mathbb{C}}_\delta(r_p, r_s)$ gives the fine horizontal displacement of
the data, $\delta$, with respect to the crude initial guess, $r_p$. The full
horizontal offset is $r_p + \delta/M$ detector pixel widths.

We originally intended to use the same approach to determine the fine vertical
offset. Unfortunately, in this case the disparity between the intrinsic
spectrum of the data and the image model strongly biases the cross-correlation
result. In a typical science image, the cutout box $\tilde{\mathbb{D}}$
contains $\sim$10 speckles. Their chromatic position dependence (as illustrated
in Figure~\ref{fig:examplecube}) causes steep brightness gradients in the
spectra formed on the spectrograph focal plane, since a given microlens will
collect light from a speckle over only a fraction of the passband. Whatever
intrinsic spectrum is built into the image model, $\mathcal{I}_M$, will
significantly differ from that of most of the sample. We found that the effects
of these disparities do not average out over an ensemble. Instead, they
systematically push the cross-correlation result up or down by a degree that
does not reflect the actual relative wavelength alignment.

Instead of using the image model as an alignment template, we return to the
fitting approach described in \S~\ref{sec:spectrummodel}. This time, however,
rather than fitting the full spectrum parameter set, we concentrate on the
region with the most information about the vertical position: the telluric
water absorption trough. Therefore, we confine the least-squares fit region to
a $3\times11$ box, aligned such that the 1.37 $\mu$m reference point is near
the middle pixel on the 8th row.

We further simplify the spectrum fit by describing the local light source with
merely two parameters: an amplitude and color. The other free parameters are
the background light offset and the vertical position. The height and tilt are
already known from the calibration image solution, and the horizontal position
is fixed based on the previous step in the registration algorithm. As in
Equation~\ref{eqn:troughcond}, the spectrum trace points with wavelengths
$1.28\:\mu\mbox{m} < \lambda < 1.52\:\mu$m are again tied to the transmission
function plotted in Figure~\ref{fig:watertrough}. The anchor points at $\lambda
= $1.28 $\mu$m and 1.58 $\mu$m are set based on the amplitude and color
parameters. 

To get a diverse set of spectrum shapes spanning a wide region of the speckle
halo, during the fine vertical offset fitting procedure we sample 121 spectra
over a $600\times600$ pixel box (as compared to the $200\times200$ pixel box
used for the horizontal registration). Of the 121 fits, the median vertical
offset is taken as the final value, and rounded to the nearest 1/7th of a pixel
to match the quantization of the fine horizontal offset. In trial runs, we
found the vertical offsets determined from the full set of sample spectra
follow a Gaussian distribution, with standard deviation 0.2--0.4 pixel widths,
depending on the source image. We accept this as the uncertainty in the
vertical registration.

\subsection{Cube Extraction}
\label{sec:cubeextraction}

\subsubsection{The Role of the Global Spectrograph Focal Plane Solution}
\label{sec:laserlookup}

In order to form a data cube, the pipeline must ``know'' where individual
spectra are positioned on the focal plane, and furthermore, which points of
those spectra correspond to a given wavelength. We rely on the global
spectrograph focal plane solution described in \S~\ref{sec:globalsol} to
establish the image geometry for each epoch under consideration. One of the
products of the calibration image fitting procedure is a text file tabulating
the positions of all $3.8\times10^4$ spectra alongside their corresponding
microlens indices. This table, combined with the results of the registration
algorithm (\S~\ref{sec:specreg}) and the maps of height and tilt parameters
give all the information needed to organize the detector data for a given
science image.

The amplitude map produced during the global fit (see
Figure~\ref{fig:param_maps}) also has an important role. It compliments the
dewar lamp flat described in \S~\ref{sec:dewarflat} by capturing the
larger-scale variations in sensitivity across the field of view. By looking up
the amplitude parameter associated with a given microlens, we can appropriately
scale any detector samples from that spectrum to compensate for optical effects
such as vignetting.

\subsubsection{Residual Background Map}
\label{sec:resbgmap}

Despite the numerous stages in the detector image processing, some minor
extraneous background structure persists into the processed focal plane image.
This component, superimposed on the real signal, is caused by a combination of
residual bias counts, scattered light within the instrument, and thermal
contamination from outside the dewar (unaccounted for in the dark subtraction).
In the cube extraction routine, after loading an individual focal plane image,
the pipeline forms a map of background count rates based on measurements
between spectra.

Figure~\ref{fig:extraction} shows the regions used to estimate the background
count rate associated with a given microlens. The upper left box is situated so
that its bottom row is matched with the $\lambda = 1.28\:\mu$m point (rounded
to the nearest row), and the bottom row of the lower right box is on level with
$\lambda = 1.67\:\mu$m. For both background boxes, the near side is spaced
three columns from the rounded center of the spectrum. 

The pipeline takes the median of the sample of the pixels in both $2\times5$
dark regions and stores this in a residual background map. After forming the
background estimates for all microlenses, the resulting map is smoothed with a
box median filter and stored for use in the inner extraction loop. 

\subsubsection{Weighted Sum Extraction}
\label{sec:weightedsum}

\begin{figure}[tb]
\epsscale{0.35}
\plotone{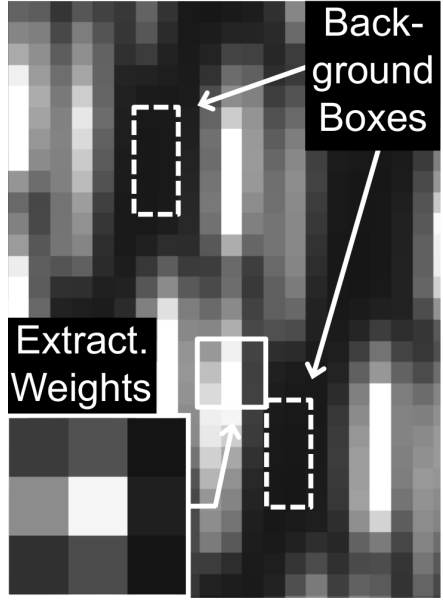}

\caption{Diagram of the P1640 cube extraction method. We step down the spectrum
in 0.03 $\mu$m increments, using the global focal plane solution to select a
$3\times3$ pixel region for each \textit{spaxel} in the cube. The sum of the
extracted samples is weighted based on the fractional offset of the true
extraction target point from the box center. In the example shown here, the
weighting scheme (bottom left inset) captures the leftward skew of the spectrum
cross section at this particular wavelength.  The two dashed boxes outline the
``dark'' regions used to sample the local background level around each
spectrum.}

\label{fig:extraction}
\end{figure}

Our cube extraction method is summarized in Figure~\ref{fig:extraction}. After
forming the background map, the extraction routine loops through microlens
indices $i$ and $j$. For each microlens, we retrieve the parameters from the
global spectrograph focal plane solution: position, height, tilt, and amplitude (represented
by the variables $(X_{0_{ij}}, Y_{0_{ij}})$, $h_{ij}$, $t_{ij}$, and $a_{ij}$,
respectively). An inner loop steps through 23 wavelength channels in 0.03
$\mu$m increments between $\lambda = 1.10\:\mu$m and 1.76 $\mu$m. We index
these channels by integer values of $w$ ($0 \le w \le 22$), and determine the
extraction target point for each cube element, or \textit{spaxel}, as follows
(cf.  Equation set~\ref{eqn:specxyparam}):

\begin{equation}
\begin{array}{lr}
X_{c_{ij}}(w) = X_{0_{ij}} + r_p + \delta/M + t_{ij}h_{ij}\left(\frac{9-w}{22}\right)\\
Y_{c_{ij}}(w) = Y_{0_{ij}} + s_p + \varepsilon/M + h_{ij}\left(\frac{9-w}{22}\right)
\end{array}
\end{equation}

\noindent where $r_p$ and $s_p$ are the crude horizontal and vertical offsets,
and $\delta$ and $\epsilon$ are the fine horizontal and vertical offset indices
determined by the registration algorithm (\S~\ref{sec:specreg}) for the current
reduced spectrograph image, $\mathbb{D}(X,Y)$.  We use a hat symbol to
designate the same coordinates rounded to the nearest pixel center:
$\left(\hat{X}_{c_{ij}}(w), \hat{Y}_{c_{ij}}(w)\right)$.

The spaxel for each microlens and wavelength combination is based on the
weighted sum over a $3\times3$ detector pixel square centered on
$\left(\hat{X}_{c_{ij}}(w), \hat{Y}_{c_{ij}}(w)\right)$: 

\begin{equation}
\mathcal{C}(i,j,w) = {\displaystyle \sum\limits_{m=-1}^1\sum\limits_{n=-1}^1 \mathbb{W}_{\alpha,\beta,w}(m,n)\left(\mathbb{D}\left(\hat{X}_{c_{ij}}(w) + m, \hat{Y}_{c_{ij}}(w) + n\right) - b_{ij}\right)/a_{ij}.}
\label{eqn:weightedsum}
\end{equation}

\noindent The weights $\mathbb{W}_{\alpha,\beta,w}(m,n)$ applied to the
detector samples are based on the PSF model, downsampled and truncated to the
$3\times3$ pixel extraction box as follows:

\begin{equation}
\mathbb{W}_{\alpha,\beta,w}(m,n) = \left\{
			\begin{array}{ll}
			{\displaystyle \Gamma_J(\alpha, \beta)\sum\limits_{i=0}^{M-1}\sum\limits_{j=0}^{M-1}\mathcal{P}_{J,M}\left(i + (m + 4)M - \alpha, j + (n + 4)M - \beta\right)\mathcal{R}_M(i,j)} \\
			\quad\quad\mbox{if}\;\; 0 \le w < 10 \\
			\:\\
			{\displaystyle \Gamma_H(\alpha, \beta)\sum\limits_{i=0}^{M-1}\sum\limits_{j=0}^{M-1}\mathcal{P}_{H,M}\left(i + (m + 4)M - \alpha, j + (n + 4)M - \beta\right)\mathcal{R}_M(i,j)}\\
			\quad\quad\mbox{if}\;\; 10 \le w \le 22 \\
			\end{array}
			\right.,
\label{eqn:weights}
\end{equation}

\noindent defined for $-1 \le m,n \le 1$. The formulas for the $J$- and
$H$-band PSFs, $\mathcal{P}_{J,M}$ and $\mathcal{P}_{H,M}$, as well as the
intra-pixel response, $\mathcal{R}_M$, can be found in \S~\ref{sec:psfmodel}.
The integers $\alpha$ and $\beta$ encode the offsets of the extraction target
point from the extraction box center:

\begin{equation}
\begin{array}{lr}
\alpha = \mathit{Round}\left(\left(X_{c_{ij}}(w) - \hat{X}_{c_{ij}}(w)\right)M\right) \\
\beta = \mathit{Round}\left(\left(Y_{c_{ij}}(w) - \hat{Y}_{c_{ij}}(w)\right)M\right).\\
\end{array}
\end{equation}

\noindent The resulting indices take on integer values in the range
$-\frac{M-1}{2} \le \alpha,\beta \le \frac{M-1}{2}$ (corresponding to offsets
up to $\frac{3}{7}$ of a pixel width in each direction when $M = 7$). Lastly,
the $\Gamma$ factor in front of each weight formula compensates for the effect
that the offset between the PSF center and the extraction box center has on the
sum of products in Equation~\ref{eqn:weightedsum}. The need for this can be
qualitatively understood by the fact that the overall flux in a $3\times3$
pixel sample of the PSF decreases when the peak is offset from the center. The
$\Gamma$ correction factor varies from unity at perfect alignment up to 1.09 in
the worst case for extreme offsets.

\subsubsection{Spectral Calibration}
\label{sec:speccal}

\begin{figure}[htb]
\epsscale{0.4}
\plotone{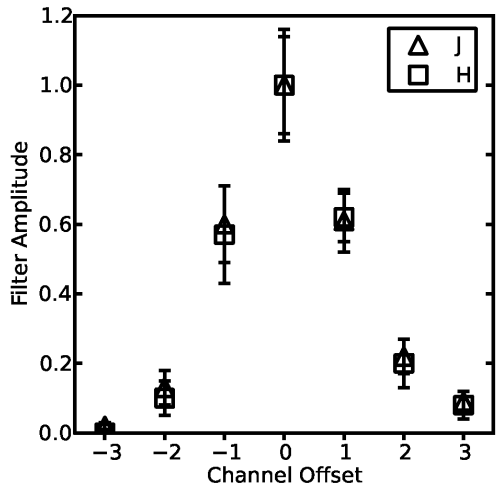}

\caption{Normalized response of the P1640 cube extraction to 1.25~$\mu$m and
1.58~$\mu$m laser sources. These are the effective filter profiles assumed for
cube channels in $J$ and $H$ band, respectively. The error bars indicate the
standard deviation of the 100$\times$100 spaxel measurement sample.}

\label{fig:laserresponse}
\end{figure}

For a given microlens, the separation between the extraction target points of
the first and last channels---corresponding to $\lambda = 1.10\:\mu\mbox{m and
}1.76\:\mu$m---is typically about 24 detector pixels. Since we use $3\times3$
pixel boxes to extract a signal for each of 23 channels spanning that length,
the footprints of adjacent channels necessarily overlap. We have examined the
effect of this by extracting data cubes directly from the laser calibration
images (discussed in \S~\ref{sec:psfmodel}). From these cubes, we compared the
mean flux in neighboring cube channels. The results, plotted in
Figure~\ref{fig:laserresponse} for 1.25 $\mu$m and 1.58 $\mu$m emission, reveal
the effective filter shape of an individual data cube channel.

The cube channel filters exhibit a full width half-maximum value of $\sim$ 70
nm at both $J$ and $H$ band. Knowledge of this profile is essential for
comparisons between P1640 data and existing astronomical spectra. We cannot
simply bin a reference spectrum to the channel spacing; we must also convolve
it with the cube channel filter before comparing it to data cube measurements.
Suppose an object appears in a data cube, and we carry out channel-wise
photometry to find a spectrum $\mathcal{A}(w)$, $0 \le w \le 22$. To compare
this meaningfully to an established spectrum, $\mathcal{B}(\lambda)$, acquired
by some other instrument with wavelength bin width $\Delta\lambda$, requires
two steps. First, we re-bin $\mathcal{B}(\lambda)$ to the cube channel
interval, 0.03 $\mu$m, to form an intermediate-resolution spectrum
$\mathcal{B}^\prime(w)$:

\begin{equation}
\mathcal{B}^\prime(w) =  \left\{
		\begin{array}{cr}
		{\displaystyle \sum\limits^{1.115 + 0.03w\:\mu\mbox{\scriptsize m}}_{\lambda = 1.085 + 0.03w\:\mu\mbox{\scriptsize m}}\mathcal{B}(\lambda)\Delta\lambda} & \quad 0 \le w \le 22 \\
		0 & \mbox{otherwise.}
		\end{array}
		\right.
\label{eqn:rebinspec}
\end{equation}

\noindent We then convolve the intermediate-resolution spectrum
$\mathcal{B}^\prime$ with the cube channel filters $\mathfrak{F}_J(z)$ and
$\mathfrak{F}_H(z)$. The filter functions are defined to follow the profiles
shown in Figure~\ref{fig:laserresponse} for $-3 \le z \le 3$ (so that $z = 0$
corresponds to the central peak of the filter), and are zero-valued outside
that range.

\begin{equation}
\mathcal{B}^{\prime\prime}(w) = \left\{
		\begin{array}{ll}
		{\displaystyle \frac{\sum\limits^{3}_{z = -3}\mathfrak{F}_J(z)}{\sum\limits^{3}_{z = -w}\mathfrak{F}_J(z)} \sum\limits^{3}_{z = -3} \mathcal{B}^\prime(w + z)\mathfrak{F}_J(z)} & 0 \le w < 10 \\
		{\displaystyle \frac{\sum\limits^{3}_{z = -3}\mathfrak{F}_H(z)}{\sum\limits^{22 - w}_{z = -3}\mathfrak{F}_H(z)} \sum\limits^{3}_{z = -3} \mathcal{B}^\prime(w + z)\mathfrak{F}_H(z)} & 10 \le w \le 22 \\
		\end{array}
		\right.
\label{eqn:convspec}
\end{equation}

\noindent The ratios in front of each convolution sum compensate for the effect
of the spectrograph passband edges on the filtering (they are unity when $w$ is
at least three channels from both passband edges). The resulting spectrum,
$\mathcal{B}^{\prime\prime}(w)$, is smoothed to the same resolution as the
cube-derived spectrum $\mathcal{A}(w)$. If $A(w)$ has been corrected for the
P1640 spectral response, then the two spectra can be directly compared, apart
from some scale factor. If on the other hand, $\mathcal{A}(w)$ is a ``raw''
cube spectrum, with undetermined spectral calibration, and
$\mathcal{B}^{\prime\prime}(w)$ refers to the same source, then it can be used
to correct $\mathcal{A}(w)$---and, in general, any P1640 data cube, as
described next.

We characterize the wavelength-dependent sensitivity of P1640, or spectral
response function, by comparing the ``raw'' data cube count values of an
unocculted reference star (observed off-axis from the coronagraph focal plane
mask) with its established near-infrared spectrum. In practice, we have chosen
stars with spectra archived in the NASA Infrared Telescope Facility (IRTF)
Spectral Library to calibrate our response~\citep{rayner}. The spectral
response function is determined by dividing the re-binned, smoothed reference
spectrum ($\mathcal{B}^{\prime\prime}(w)$, in the notation above) by the
spectrum of the same source derived from a P1640 data cube.

\begin{figure}[htb]
\epsscale{0.45}
\plotone{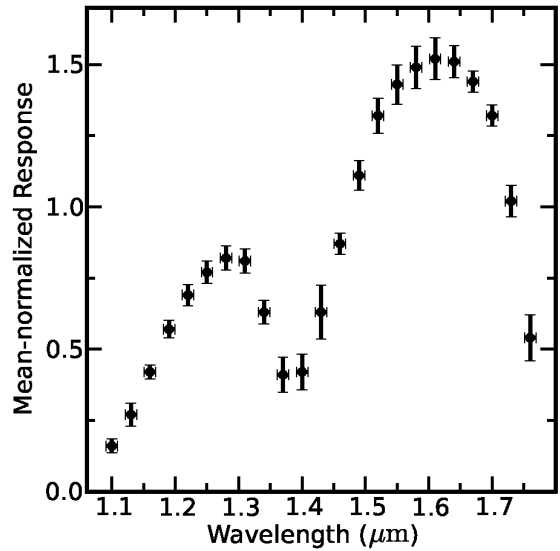}

\caption{P1640 response curve, determined through a comparison between
photometry on an unocculted star's data cube (HD 75555, in this case) with the
established spectrum of the star archived in the Infrared Telescope Facility
Spectral Library.}

\label{fig:response}
\end{figure}

We measure the signal of the observed reference star by carrying out aperture
photometry on each channel image making up the data cube, enclosing the third
Airy ring. To capture the wavelength-dependent scaling of the coronagraph PSF,
we linearly increased the photometric aperture radius from 13 to 20 spaxels
across the passband. The resulting response curve for one calibration star, HD
75555 ($V = 8.1$; spectral type F5.5III-IV), is shown in
Figure~\ref{fig:response}. The response curve shows the expected roll-off at
the edges of the operating range due to telluric water absorption features. The
valley centered near $1.4~\mu$m is likewise due to water absorption between $J$
and $H$ band. The overall climb in the response towards longer wavelengths is
caused by the wavelength-dependence of three effects in combination: the energy
per photon as dictated by the Plank relation, $E = \frac{hc}{\lambda}$;
detector quantum efficiency; and the transmission of the blocking filter at the
IFU entrace.

We normalize the spectral response function to its mean value before storing it
for general application to data cubes. The cube extraction pipeline loads one
of these mean-normalized spectral response functions into memory before
beginning the cube extraction routine. If the appropriate switch is set by the
user, the pipeline will divide the spaxel values in each channel image by the
corresponding response function value. 

\subsubsection{Error Sources}

Uncertainty in the global spectrograph focal plane solution, the image
registration, and the weighting function combine to contribute a pseudo-random
error to each cube point, at a level of $\sim$ 5\% of the spaxel value. We
extracted cubes from the Moon calibration images to estimate the magnitude of
this error. Since spatial amplitude variations are compensated for during
extraction, ideally any given spatial cross-section of a cube extracted from a
calibration image would appear flat. Instead, we observe a standard deviation
of 3\%, with some variation between channels and areas of the image.

\begin{figure}[htb]
\epsscale{0.6}
\plotone{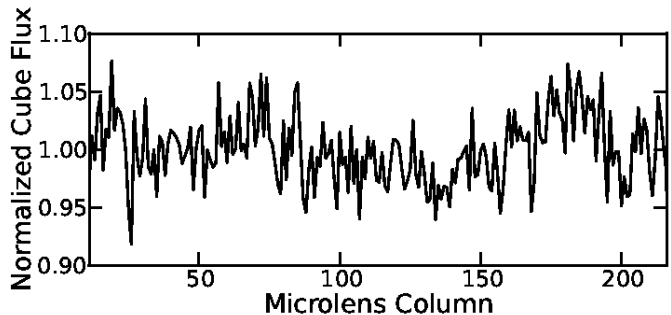}

\caption{A row in a data cube extracted from a June 2009 Moon calibration
image, with flux values normalized to unity. The scatter in flux across the row
(here the standard deviation is 3\%) reveals flux-proportional errors in the
cube extraction.}

\label{fig:flatnoise}
\end{figure}

There are also systematic errors caused by light from adjacent microlenses
overlapping on the focal plane, which we refer to as cross-talk.  The
horizontal space between spectra on the focal plane is 3.3 detector pixels, and
yet we know from the spectrograph PSF model
(Figure~\ref{fig:psf_cross_sections}) that about 10\% of the downsampled PSF
flux falls outside the central three columns. As a result, a small fraction of
light from one microlens is inevitably counted during the extraction of a
neighboring spectrum. Consider the spectrograph image cutout shown in
Figure~\ref{fig:extraction}. For each channel of a given spectrum, you can
attribute the dominant contamination to a different channel belonging to the
neighboring spectrum positioned either above or below along the microlens
column. We know from examining the laser calibration data cubes that the upper
limit of the flux incorrectly extracted into a cube point is $\sim$ 5\% of the
cube value in a neighboring spectrum. In other words, suppose
$\mathcal{C}(i_0,j_0,w_0)$ is the cube point whose contamination we are trying
to assess. Based on our spectrum model, we can determine that some channel
$w_1$ of microlens $(i_0,j_0-1)$ is the dominant contamination source for
channel $w_0$ of microlens $(i_0,j_0)$. Therefore, we estimate the cross-talk
error $0.05\mathcal{C}(i_0,j_0-1,w_1)$, added in quadrature with the
uncertainty described earlier. Using a table of established cross-talk channel
pairs, we can repeat this error estimate for each channel of a spectrum of
interest. The spectral response function plotted in Figure~\ref{fig:response}
and the source spectrum plotted in Figure~\ref{fig:titan} reflect this
analysis.

\subsection{Pipeline Data Products}

The PCXP stores the reduced data output in FITS files, and organizes them in a
directory tree by object and date. Three channels from an example cube based on
an occulted star observation are displayed in Figure~\ref{fig:examplecube}. In
addition to the normal cube extraction described in
\S~\ref{sec:cubeextraction}, there are several other products the pipeline
derives from the raw data. From the brightest stars, there is enough signal
recorded in a single 7.7 second read to form a cube without using the full
exposure time. One option of the pipeline takes advantage of this, checking if
the $V$-band magnitude is less than 2.0, and if so then making cubes from each
pair of consecutive reads in the NDR sequence.  The resulting ``read-wise''
cubes have a speckle pattern resolved to a higher time resolution, which may
eventually be exploited to improve speckle suppression. At the opposite time
scale, the pipeline can form cubes from the mean of all spectrograph images
acquired on the same data of a given target. The pipeline also forms
``collapsed'' images by summing all the channel images of a cube, as well as
the subsets of channels corresponding to the $J$ and $H$ bands.

\section{Example Spectrum Retrieval: Titan}
\label{sec:titan}

\begin{figure}[htb]
\epsscale{0.45}
\plotone{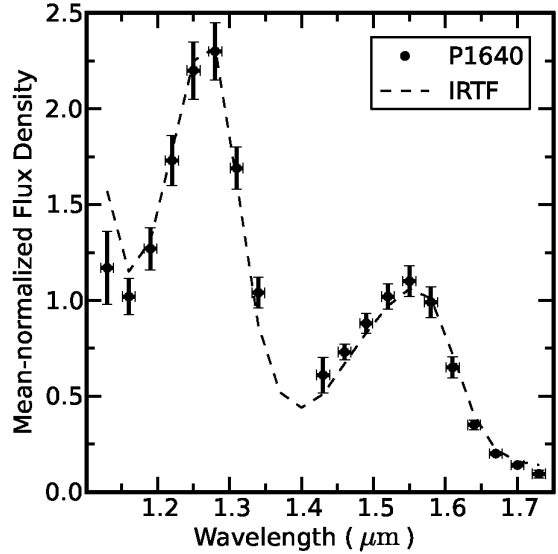}

\caption{Disk-averaged spectrum of Titan extracted from a P1640 data cube,
calibrated using the spectral response function in Figure~\ref{fig:response}.
For comparison, a near-infrared spectrum of Titan acquired with IRTF/SpeX by
Schaller is plotted alongside the P1640 data, after being binned and smoothed
to the P1640 cube resolution. The IRTF/SpeX and P1640 Titan data were acquired
two days apart, on 2009 March 13 and 15, respectively. Each spectrum is
normalized to the mean of the points included in the plot. Channels 1, 10, 11,
and 23 ($\lambda = 1.10\:\mu\mbox{m}, 1.37\:\mu\mbox{m}, 1.40\:\mu\mbox{m, and
}1.76\:\mu\mbox{m}$, respectively) were excluded due to varying telluric water
absorption.}

\label{fig:titan}
\end{figure}

To demonstrate the efficacy of our data extraction and calibration procedures,
we apply them here to an observation of Saturn's moon Titan acquired on 2009
March 15. After locking the AO system on Titan, its image was positioned
off-axis from the focal plane mask so that no part of the $1\arcsec$ disk was
occulted by the coronagraph. We used the pipeline to generate a data cube from
a single 138 s exposure, calibrating the relative channel fluxes with the
response function shown in Figure~\ref{fig:response}. Next, we averaged 1900
spaxels inside the resolved disk of Titan. After normalizing the disk-averaged
spectrum to the mean channel flux, we compared it to unpublished data obtained
by Emily Schaller two days earlier using the SpeX near-infrared
spectrograph~\citep{spex} at the NASA Infrared Telescope Facility (IRTF).
Following the procedure in~\S~\ref{sec:speccal}, we re-binned and smoothed the
IRTF spectrum to match the resolution of the data cube
(Equations~\ref{eqn:rebinspec}--\ref{eqn:convspec}). The resulting spectra are
plotted in Figure~\ref{fig:titan}. The near-infrared spectrum of the moon is
marked by a series of broad CH$_4$ absorption troughs~\citep{fink}. At the two
albedo peaks in our passband, 1.3$~\mu$m and 1.6$~\mu$m, Titan's atmospheric
opacity is low enough for the direct reflection of sunlight off the water ice
surface to constitute the observed flux, rather than diffuse scattering in its
stratospheric haze~\citep{griffith}.

We expect the disk-averaged spectra acquired on these two dates to be similar.
As Titan rotates over a 16-day period, in synchronicity with its orbit around
Saturn, the near-infrared albedo observed from Earth (through the 1.3$~\mu$m
and 1.6$~\mu$m methane ``windows'') varies in a cycle with an amplitude on the
order of 10\%.  This variation is caused by a change in surface features
between the leading and trailing hemispheres~\citep{lemmon}.  However, despite
the 45 degree rotation of Titan with respect to Earth between the IRTF and
P1640 observations, previous monitoring by several investigators indicates no
significant albedo change between our specific pair of planetographic
longitudes ($197\degr$ and $243\degr$)~\citep{griffith1998}.  Furthermore,
long-term monitoring of Titan's albedo only occasionally reveals deviations
from predicted reflectivity due to transient cloud
features~\citep[e.g.][]{griffith1998,schaller}. The anticipated resemblance of
the two spectra is confirmed in Figure~\ref{fig:titan}: the average absolute
difference between the IRTF and P1640 Titan data over the 19 channels used in
Figure~\ref{fig:titan} is 7\% of the mean flux, and most flux points agree
within the error bars of the P1640 data. The only channel flux showing
significant disparity with the IRTF data, centered at 1.13 $\mu$m, is located
near the edge of a telluric water absorption trough, and therefore is more
susceptible to calibration errors than most channels on the plot. For examples
of M dwarf stellar spectra that have been measured from P1640 data cubes,
see~\cite{alcor} and~\cite{zetavir}.

\section{Discussion}
\label{sec:discussion}

Starting from the economic constraint of limited detector area, the design of
any integral field spectrograph must reach compromises between the competing
parameters of spatial resolution, field of view, spectral resolution, and
spectral range. For Project 1640, the need to Nyquist-sample the starlight
speckle pattern inside the angular extent of the adaptive optics system
``control radius'' largely determined the balance of these tradeoffs. The other
major factor was the overarching science goal of distinguishing astrophysically
interesting features in the spectral energy distributions of young giant
exoplanet atmospheres. Based on these considerations, the P1640 collaboration
concluded on an IFU design that strongly favored a high density of spatial
elements over spectral resolution, to a greater extreme than previous
microlens-based integral field spectrographs. For example, the broadband mode
of the TIGER IFU had 572 spatial elements dispersed at spectral resolution
$R\:\sim\:370$, and the broadband mode of the OSIRIS IFU has 1024 spatial
elements dispersed at $R\:\sim\:3400$~\citep{bacon1995, larkin2006}.  P1640, by
comparison, has 38,000 spatial elements with $R\:\sim\:40$, corresponding to
over a factor of 30 increase in spatial elements, and a similarly substantial
reduction in spectral resolution. Only two other IFUs will join P1640 in this
operating regime in the next year: GPI and VLT-SPHERE---both designed for
exoplanet imaging.

Since the properties of P1640 data are unusual even in the context of preceding
IFUs, the instrument commissioning has required development of novel extraction
and calibration approaches. Consider that each detector pixel width spans
approximately 27.5 nm in spectral dispersion---4\% of the instrument
passband---and that the physical length of an individual microlens spectrum's
footprint is merely 0.5 mm at the focal plane. As such, great care has been
required to accurately map the spectrograph focal plane data, in a manner that
is resilient to subtle long-term changes in optical alignment. We described an
answer to this problem in \S~\ref{sec:focplmodel}, using a hierarchical fitting
procedure to build a comprehensive, epoch-specific model of the full
spectrograph focal plane. The last obstacle to securing the layout of the data,
instrument flexure, must be dealt with individually for each exposure.
Therefore, unlike the case of building the spectrograph focal plane model, we
register each science spectrograph image ``on the fly'' inside the data
pipeline, as explained in \S~\ref{sec:specreg}.

The relatively long execution time required to build the spectrograph focal
plane solution ($\sim$ 12 hours on a single high-performance workstation) means
that it is sometimes necessary to rely on an outdated solution to extract data.
Such will be the case, for example, during the first night of an observing run
when the calibration Moon/sky image has not yet been acquired and fit. We know
from the geometric evolution illustrated in Figure~\ref{fig:distortion} that
consequent errors in the positions of the spectra, along with other properties,
will inevitably degrade the quality of the cube.  However, for a preliminary
inspection of data, and to check on instrument performance, the result will
usually be acceptable. Recall that the spectrograph image registration uses a
region of the image with the strongest signal to align the model. Therefore,
under circumstances where the solution is old, the extraction will still tend
to be fairly accurate near the brightest region of the image, but will
progressively worsen towards the outskirts of the focal plane.

The main advantages of our weighted sum approach to IFU spectrum extraction,
described in \S~\ref{sec:weightedsum}, are simplicity and speed. Our reasoning
behind using the spectrograph PSF itself to shape the weighting function is
that it mimics the cross-dispersion profile of the spectrum. This way, along
the middle horizontal row of the $3\times3$ extraction box, we weight the
detector samples by an estimate of their relative intensity (see
Figure~\ref{fig:extraction} for an example).~\cite{horne} originally devised
this strategy for extracting coarsely-sampled CCD spectra. He demonstrated that
matching the weights to the expected cross-dispersion profile optimizes the
fidelity of the extraction in the case where read noise is prevalent. This
makes sense intuitively because we want detector samples with high count rates
to have more influence than weak, noisy ones. Unlike the case of Horne's
spectrograph, however, we are dealing with many closely packed spectra, and so
we are forced to truncate our weighting area to a region smaller than the
actual extent of the PSF. We attempt to account for this in the weighting
formula in Equation~\ref{eqn:weights}. Since this correction depends not just
on the shape of the PSF but also the target spectrum, it is necessarily only an
estimate, and can contribute errors on the order of a few percent to the spaxel
value.

As compared to the cross-dispersion axis, for the dispersion axis there is more
freedom in the choice of extraction weights. The tradeoffs here are
signal-to-noise ratio per channel versus spectral resolution. Ultimately, for
any weighted sum approach, the spectral resolution is limited by the width of
the monochromatic IFU response along the dispersion axis. The spectrograph PSF
fitting result plotted in the right-hand panel of
Figure~\ref{fig:psf_cross_sections} shows this is about 2 detector pixels for
P1640, which translates to $\sim$ 50 nm in the dispersion direction. For
simplicity, we chose to remain with the PSF again to set the weights; the
resulting cube spectra have a resolution of about 70 nm ($R\:\sim\:20$). In the
future, it would be worthwhile to experiment with a hybrid weighting function
that combines the cross-dispersion profile of the PSF with a different vertical
profile, to investigate how much the spectral resolution can be improved. As
one example of an alternative,~\cite{maire2010} propose extracting spectra from
the GPI IFU by summing strictly along a single row/column of pixels in the
cross-dispersion direction. 

Another possible improvement to our data pipeline is a completely different
extraction approach based on deconvolution or fitting. A well-designed fitting
algorithm might disentangle the flux contributions of wavelengths with
overlapping footprints. We have done a few experiments in this direction---for
example, we applied the same MPFIT-based algorithm we used to model the
microlens spectra in the Moon/sky calibration image
(\S~\ref{sec:spectrummodel}) to a science image. The results were of
significantly poorer quality than our normal data cubes, partly as a
consequence of not having the luxury to average the fit spectra over many
microlenses, as we do with calibration images. In another program, two of our
co-authors created a program that fits each spectrum cutout as a train of
scaled PSFs, each one representing a different channel. While its
implementation is not complete, this method shows promise of forming data cubes
with slightly higher spectral resolution than the existing extraction. Since
any fitting approach is inherently much slower than a weighted sum translation,
one can imagine two cube extraction algorithms co-existing for different
purposes: one as an offline procedure reserved for images of the greatest
interest, and the other program applied to all P1640 data. 

\section{Conclusions}

We have developed a collection of algorithms to reduce the data acquired by
P1640, a coronagraphic integral field spectrograph designed for high contrast
imaging. Our aim has been to describe our data pipeline software in enough
detail that upcoming microlens-based imaging spectrograph projects can take
advantage of our experience in treating closely packed, coarsely sampled
spectra.

An essential element of our approach is an empirical model of the spectrograph
focal plane image, based on calibration exposures in which the entire microlens
array is illuminated, in turn, by broadband and monochromatic light. To derive
a solution specific to each observation epoch, we fit a set of parameters
describing each microlens spectrum (position, tilt, height, and overall signal
amplitude). We use the resulting table of solved parameters to determine the
extraction location on the focal plane for any given combination of microlens
and wavelength, and hence build the data cube. 

We implement the cube extraction with a weighted sum that optimizes the
signal-to-noise ratio by mimicking the expected cross-dispersion profile, as
constrained by the sub-pixel spectrograph image registration. Sources of error
in the final data cube are cross-talk between adjacent microlens spectra,
uncertainty in the spectrograph focal plane model, uncertainty in sub-pixel
registration, and uncertainty in the determination of extraction weights.
Nevertheless, based on an observation of Saturn's moon Titan, we have
demonstrated our ability to retrieve strong-featured near-infrared spectra to
$\sim$ 5\% accuracy. As our methods for handling this new form of data evolve,
we expect P1640 to continue its pioneering role in high contrast astronomy.

\acknowledgments

Thanks are due to Anand Sivaramakrishnan, Emily Rice, Michael McElwain, James
Gunn, David Zurek, and Charles Beichman for helpful discussions. We also thank
Emily Schaller for providing us with unpublished SpeX/IRTF Titan data to serve
as a reference spectrum in Figure~\ref{fig:titan}. Project 1640 is funded by
National Science Foundation Grants AST-0520822, AST-0804417, and AST-0908484.
Part of this work was performed under a contract with the California Institute
of Technology (Caltech) funded by NASA through the Sagan Fellowship Program.
The members of the Project 1640 team are also grateful for support from the
Cordelia Corporation, Hilary and Ethel Lipsitz, the Vincent Astor Fund, Judy
Vale, Andrew Goodwin, and an anonymous donor.

{\it Facilities:} \facility{Hale (PALAO, Project 1640)}

\end{document}